\documentclass[journal]{IEEEtran}
\usepackage{anyfontsize}
\usepackage{orcidlink}
\usepackage{cite}
\usepackage{amsmath,amssymb,amsfonts}
\usepackage{algorithm}
\usepackage{algpseudocode}
\usepackage{graphicx}
\usepackage{textcomp}
\usepackage{xcolor}
\usepackage{import}
\usepackage{amsmath}
\usepackage{array}
\usepackage{booktabs}
\usepackage{multirow}
\usepackage{caption}
\usepackage{hyperref}
\usepackage{amssymb}
\usepackage{pifont}
\usepackage{threeparttable}
\usepackage{subcaption} 
\usepackage{float} 
\usepackage[export]{adjustbox}

\newcommand{\cmark}{\textcolor{teal}{\ding{51}}}
\newcommand{\xmark}{\textcolor{purple}{\ding{55}}}

\ifCLASSINFOpdf
\else
\fi
\begin{document}
%
\title{\fontsize{22}{25}\selectfont MOFCO: \underline{M}obility- and Migration-Aware Task \underline{O}ffloading in Three-Layer \underline{F}og \underline{Co}mputing Environments}
%
%
%

\author{ \IEEEauthorblockN{Soheil~Mahdizadeh\orcidlink{0009-0007-0065-8098},
        Elyas~Oustad\orcidlink{0009-0006-1456-356},
       and~Mohsen~Ansari\orcidlink{0000-0002-4670-8608}}
\thanks{Manuscript received 
     15 July 2025; revised MM DD, 2025 and MM DD, 2025; accepted MM DD, 2025. Date of publication MM DD, YY; date of current version MM DD, YY. \textit {(Corresponding author: Mohsen Ansari.)}
}
\thanks{The authors are with the Department of Computer Science and  Engineering, Sharif University of Technology, Tehran, Iran, (e-mail: \{s.mahdizadeh, e.oustad, ansari\}@sharif.edu).}
}

\maketitle

\begin{abstract}
Task offloading in three-layer fog computing environments faces a critical challenge with user equipment (UE) mobility, which frequently triggers costly service migrations and degrades overall system performance. This paper addresses this problem by proposing MOFCO, a novel Mobility- and Migration-aware Task Offloading algorithm for Fog Computing environments. The proposed method formulates task offloading and resource allocation as a Mixed-Integer Nonlinear Programming (MINLP) problem and employs a heuristic-aided evolutionary game theory approach to solve it efficiently. To evaluate MOFCO, we simulate mobile users using SUMO, providing realistic mobility patterns. Experimental results show that MOFCO reduces system cost—defined as a combination of latency and energy consumption—by an average of 19\% and up to 43\% in certain scenarios compared to state-of-the-art methods.

\end{abstract}


%
\IEEEpeerreviewmaketitle

\section{Introduction}

\IEEEPARstart{W}{ith} the rapid advancement of mobile communication technologies, new services such as the Internet of Things (IoT) have emerged, followed by the Internet of Vehicles (IoV), both demanding increased resources from current wireless systems, including throughput, bandwidth capacity, and computation capacity \cite{b1}. To meet these demands, the fifth generation (5G) wireless systems were developed \cite{b2}. Initially, tasks were processed on local computational devices, or user equipment (UEs), often constrained by limited computational power and battery lifetime. Cloud radio access networks (C-RANs) reduced these issues by offloading tasks to centralized data centers, thus lowering local energy consumption and using substantial computational resources \cite{b1}. However, the substantial distance between UEs and cloud data centers often resulted in high transmission latency, making it difficult for latency-sensitive tasks to be executed properly \cite{b2}.

To address the latency issues inherent in cloud computing, fog radio access networks (F-RANs) were introduced \cite{b2}. Fog computing acts as a middle layer between the UEs and C-RANs. It brings computational resources closer to the end users, thus reducing latency by processing data locally at these fog nodes, which is crucial for real-time applications that require prompt processing \cite{b3}. Examples include traffic management systems that dynamically control signals to prevent congestion, autonomous vehicle networks that require near-instantaneous communication between vehicles and infrastructure, and emergency response systems that monitor environmental data for early detection of hazards. In smart healthcare, real-time patient monitoring systems rely on fog nodes for quick processing of critical health metrics, ensuring timely medical intervention \cite{b4}. Due to the improved power efficiency of these computation nodes, the overall energy that the tasks impose on battery-sensitive devices can also be reduced.

\begin{figure}[t]
\centerline{\includegraphics[width=\columnwidth]{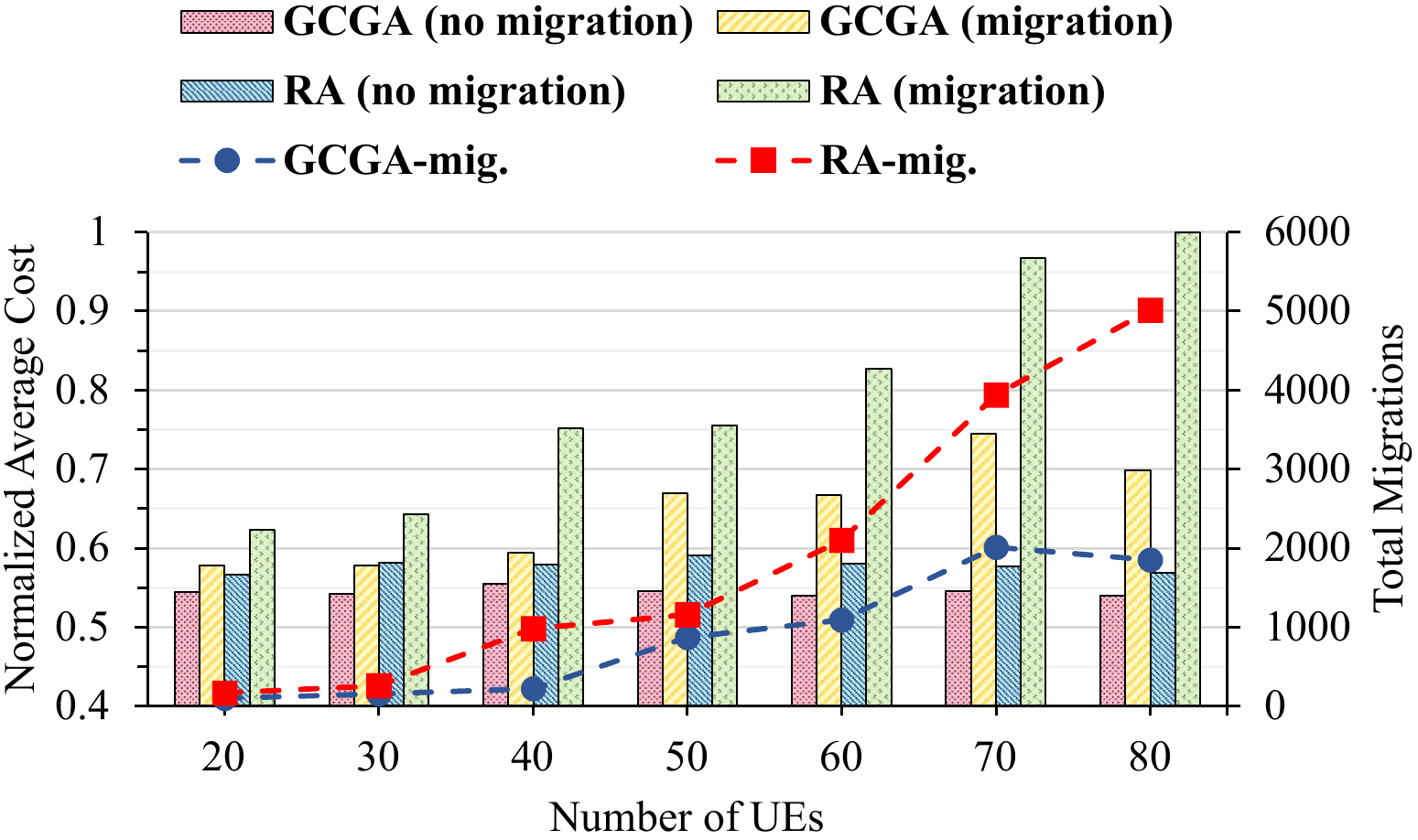}}
\caption{Total cost of baseline methods with and without considering the extra cost of migration.
}
\label{observationFig}
\vspace{-1.75em}
\end{figure}

The main purpose of a three-layered fog computing architecture is efficiently offloading user equipment (UE) tasks to the fog layer, as selecting the appropriate fog node significantly impacts overall system latency and load balancing. Numerous studies have proposed various methods to address this issue \cite{b4, b5, b6, b7, b8}. Works in \cite{b14, b15, b16, b17, b18, b19, b20} have also considered the joint problem of offloading decisions and resource allocation. Despite these advancements, user equipment (UE) mobility introduces new complexities such as radio handover and service migration \cite{b9, b10}. Due to mobility, UEs might leave the coverage of a fog node, resulting in additional energy consumption and time delay. Migration failure may also occur due to late migration, early migration, or migration to a fog node with much higher latency. Additionally, the target fog node may not have enough available capacity to support new UEs, causing the newly migrated tasks to wait in line for execution, increasing time delay, and occupying storage space. To show the impact of migration, we ran a preliminary simulation that highlights its significant cost.

\subsection{Observation}

To further illustrate the critical role of migration and its impact on highly mobile environments, we simulated a scenario using SUMO \cite{b11} and the results are depicted in Fig. \ref{observationFig}. A detailed explanation of the simulation environment and system model will be provided in subsequent sections; however, for now, consider that the GCGA and RA baseline methods were simulated. Further explanation of these methods is available in section \ref{simulationResults}, but for now, note that GCGA prioritizes fog nodes, and RA is a random method, randomly choosing the offloading destinations. First, we simulated these methods in a scenario without considering migration (in this case, if a mobile UE exits the coverage area of its destination fog node, no extra cost is calculated). Then, we added the extra cost of migration, and as seen in the figure, an increasing number of UEs results in more tasks, and the total number of migrations also increases. The difference in cost between the two scenarios is apparent, and the cost that migration imposes on the system must not be neglected, especially in highly populated environments where it can make a significant difference in the final cost.

When accounting for UE mobility, offloading decisions must consider a wider range of factors, including the remaining computational resources at both the local and higher-level nodes, as well as the duration the UE remains within the coverage area of proximate fog nodes \cite{b7}. To further illustrate how parameters like available resources and mobility might affect the offloading decision, the following motivational example is presented.

 

\subsection{Motivational Example}
\label{motivationalexample}

\begin{figure}[t]
\centerline{\includegraphics[width=\columnwidth]{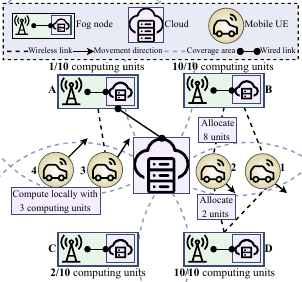}}
\caption{Illustrative example of various scenarios occurring in a mobile three-layer network topology
}
\label{motivationalFig}
\vspace{-1.75em}
\end{figure}

In this section, we present four scenarios within the three-layer topology, highlighting how the mobility of UEs impacts task offloading decisions, the selection of offloading destinations, and resource allocation strategies. \hyperref[motivationalFig]{Fig. 2} illustrates four scenarios, which are further explained in the following paragraphs.

\begin{itemize}
    \item \textbf{Scenario 1:}  
    In this scenario, without considering a specific resource allocation after offloading, the first UE is within the coverage area of two fog nodes, B and D, but is moving closer to the coverage area of fog node D. Offloading to fog node D appears more rational, as offloading to fog node B could result in the task's completion when the UE has already exited its coverage area, leading to migration. Offloading to fog node D, however, offers a higher likelihood of avoiding migration since the UE is moving toward it. If the task is completed within a certain timeframe, the result can be returned to the UE without the overhead of migration.

    \item \textbf{Scenario 2:}  
    Similar to the first scenario, the second UE is also within the coverage of fog nodes B and D. However, this time, we consider the resource allocation policy: offloading to fog node B provides 8 computing units while offloading to fog node D provides only 2 computing units. The execution time of the offloaded task varies between these two strategies. Offloading to fog node D increases the risk of migration, as the lower computing power could result in a longer execution time, causing the UE to exit the coverage area before completion. Although the UE is moving away from fog node B's coverage area, offloading it with 8 computing units significantly reduces the execution time, thereby decreasing the probability of migration in this scenario.

    \item \textbf{Scenario 3:}  
    In this scenario, the third UE is within the coverage of fog nodes A and C. However, due to prior resource allocations for other tasks, both fog nodes have limited computing resources available. Offloading to either fog node is not ideal, as the limited computing power may result in an execution time long enough for the UE to exit both coverage areas. In this case, offloading to the cloud layer could be advantageous, as it reduces energy consumption and completely eliminates the risk of migration, though it might increase system latency. To execute this offloading mechanism, the task must first be offloaded to the nearby fog node via a wireless link and then transmitted to the cloud via a wired link.

    \item \textbf{Scenario 4:}  
    Similar to the third scenario, this UE is moving in the same direction and is within the coverage of fog nodes A and C, which have limited computing resources available. Another option is to avoid offloading and compute locally. This approach eliminates the probability of migration but increases energy consumption. Nevertheless, as with the cloud offloading scenario, this could be beneficial due to the absence of task migration.
\end{itemize}

In these four scenarios, we have discussed the challenges and complexities that mobility introduces to a fog computing topology. We have also demonstrated that each of the three layers in the architecture offers distinct advantages, with trade-offs between energy consumption and latency.

\subsection{Contributions}

While numerous studies have explored task offloading strategies to reduce energy consumption and improve latency \cite{b4, b5, b6, b7, b12, b13, b14, b15, b16, b17, b18, b19, b20}, mobility has often been inadequately addressed. Our work distinguishes itself by integrating a real-time network topology model that includes the geographical positions and movement dynamics of UEs and fog nodes. It also models the migration of tasks and accounts for the additional cost it introduces to the system. Although some studies have focused on reducing the cost of migration itself \cite{b9, b10}, our emphasis is on reducing its occurrence, as the cost of migration cannot be completely eliminated and will, more or less, cause extra costs, especially in high-demand situations. This model enables task-offloading decisions based on the current physical environment rather than probabilistic approaches. As a result, our proposed algorithm provides solutions by accurately reflecting dynamically changing network topologies. The main contributions of this work can be summarized as follows:

\begin{figure}
\centerline{\includegraphics[width=\columnwidth]{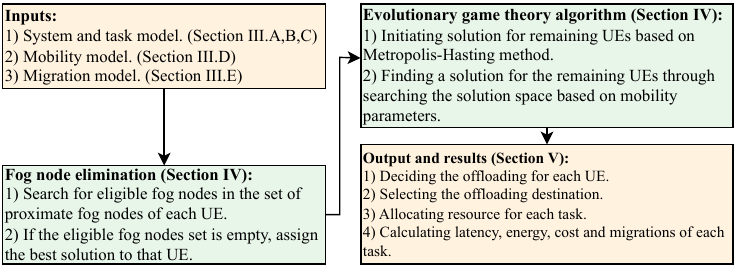}}
\caption{Overall structure of our paper}
\label{structureFig}
\vspace{-1.75em}
\end{figure}

\begin{enumerate}
    \item \textbf{Mobility- and Migration- Aware Fog Computing Model and Optimization:} In Section~\ref{systemmodel}, we introduced a realistic three-layer fog architecture—comprising local, fog, and cloud layers—designed to support mobile environments where the proximity of fog nodes to UEs changes dynamically based on location and velocity. The model allows continuous task generation and supports task migration through the cloud with associated costs. Building on this, in Section~\ref{problemFormulationAndSolution}, we proposed MOFCO, a heuristic-aided evolutionary game theory algorithm that solves the offloading decision, destination selection, and resource allocation problem. The optimization is formulated as a Mixed-Integer Nonlinear Programming (MINLP) problem and aims to minimize total task cost, considering latency, energy usage, and migration overhead.

    \item \textbf{Future Workload Prediction Using Mobility Parameters:} Unlike existing approaches, our method utilizes mobility parameters to predict future workload distribution. This predictive foresight, derived from real-time UE trajectories, serves as the core mechanism within MOFCO that enables a strategic shift from reactive migration handling to proactive migration avoidance, thereby enhancing decision making under dynamic conditions.
    
    \item \textbf{Performance Evaluation in Realistic Environments:} Finally, in section \ref{simulationResults}, the proposed method is simulated using SUMO, a well-known simulation tool for modeling mobile environments. The results demonstrate migration-avoiding performance, with the scenarios tested based on real-world conditions over an extended period during high-traffic situations. Experimental results confirm that MOFCO achieves higher cost-efficiency than existing state-of-the-art solutions.
    
\end{enumerate}

The overall structure of the paper is shown in Fig. \ref{structureFig} and is as follows: Section II reviews related work and discusses proposed methods and their associated issues. Section III presents our proposed system model and explains its components. Section IV details
the proposed evolutionary game theory algorithm. Finally, Section V evaluates the performance of the proposed method through various simulations and compares it with baseline methods to validate its effectiveness.

\section{Related Works}
\label{relatedworks}
In this section, we have organized the related works into three subsections, highlighting all aspects of the problem.

\subsection{Task Offloading and Resource Allocation in Fog Computing Networks}

The primary focus of all related works revolves around the offloading problem and resource allocation. In studies such as \cite{b12}, the author proposes a game-theoretic approach to address the offloading problem. In \cite{b13}, the authors explore partial offloading in fog networks, comparing centralized and distributed architectures to balance energy consumption and task delay. The work in \cite{b14} addresses a fairness-aware cost minimization problem for user equipment (UE), differing from previous studies by focusing on minimizing the maximum cost among all UEs. In \cite{b15}, the FAJORA framework is presented for joint optimization, addressing offloading decisions, resource allocation, uplink pattern, and power allocation. Similarly, the work in \cite{b16} introduces a QoE maximization framework for IoT users' computation offloading decisions, modeled as a potential game with a proven pure Nash equilibrium.

Although all these works focus on offloading decisions, destination selection, and resource allocation, none have considered a mobile environment or incorporated a mobility model for user equipment.

\begin{table*}[ht]
\centering
\caption{Comparison of Related Works}
\begin{threeparttable}
\resizebox{\textwidth}{!}{
\begin{tabular}{ccccccccc}
\hline
\textbf{Reference} & \textbf{Local Execution} & \textbf{Fog Layer} & \textbf{Cloud Layer} & \textbf{Resource Allocation} & \textbf{Mobility} & \textbf{Migration} & \textbf{Mobility Model} & \textbf{Optimized Parameter} \\ 
\hline
\cite{b12} & \cmark & \cmark & \cmark & \xmark & \xmark & \xmark & \xmark & Latency \& cost \\ \hline
\cite{b13} & \cmark & \cmark & \xmark & \xmark & \xmark & \xmark & \xmark & Latency \& energy \\ \hline
\cite{b14} & \cmark & \cmark & \cmark & \cmark & \xmark & \xmark & \xmark & Latency \& energy \\ \hline
\cite{b15} & \cmark & \cmark & \cmark & \cmark & \xmark & \xmark & \xmark & Resource utilization \\ \hline
\cite{b16} & \cmark & \cmark & \cmark & \cmark & \xmark & \xmark & \xmark & Latency \& QoS \\ \hline
\cite{b17} & \cmark & \cmark & \xmark & \cmark & \xmark & \xmark & \xmark & Profit \& delay \\ \hline
\cite{b18} & \xmark & \cmark & \xmark & \cmark & \xmark & \xmark & \xmark & Capacity of links \\ \hline
\cite{b19} & \cmark & \cmark & \xmark & \cmark & \xmark & \xmark & \xmark & Latency \& energy \\ \hline
\cite{b20} & \cmark & \cmark & \xmark & \cmark & \xmark & \xmark & \xmark & Latency \& energy \\ \hline
\cite{b21} & \xmark & \cmark & \xmark & \xmark & \cmark & \xmark & Access prediction & Energy \& success rate \\ \hline
\cite{b22} & \xmark & \cmark & \cmark & \cmark & \cmark & \xmark & Mobility pattern & Service satisfaction \\ \hline
\cite{b23} & \cmark & \cmark & \xmark & \xmark & \cmark & \xmark & Location prediction & Delay \& success rate \\ \hline
\cite{b24} & \cmark & \cmark & \xmark & \xmark & \cmark & \xmark & Straight highway & Delay \\ \hline
\cite{b25} & \cmark & \cmark & \xmark & \xmark & \cmark & \xmark & Straight highway & Delay \\ \hline
\cite{b26} & \cmark & \cmark & \cmark & \xmark & \cmark & \xmark & Straight highway & Delay \& cost \\ \hline
\cite{b27} & \cmark & \cmark & \xmark & \xmark & \cmark & \xmark & SUMO location & Average response time \\ \hline
\cite{b28} & \cmark & \cmark & \xmark & \xmark & \cmark & \cmark & SUMO location & Latency \\ \hline
\cite{b29} & \cmark & \cmark & \xmark & \cmark & \cmark & \cmark & Sojourn time & Latency \& energy \\ \hline
Our MOFCO method & \cmark & \cmark & \cmark & \cmark & \cmark & \cmark & SUMO location \& velocity & Latency \& energy \\ 
\hline
\end{tabular}
}
\end{threeparttable}
\vspace{0.1cm} \\
(The parameters that have been taken into account are marked with a \cmark \:symbol, while those not considered are indicated by a \xmark \:symbol.)
\label{table:related_works}
\vspace{-1.75em}
\end{table*}

\subsection{Mobility-aware in Fog Computing Networks}
In the following previous works, the main problems of fog computing networks in a mobile environment have been addressed, with some considering actual mobile user equipment in their proposed models.

In \cite{b17}, a MEC-enabled blockchain framework for video streaming is introduced, featuring an incentive mechanism to promote collaboration among content creators, transcoders, and consumers without third-party intervention. The key contributions of \cite{b18} include (1) proposing a decentralized resource allocation mechanism using deep reinforcement learning (DRL) to address latency constraints on V2V links, and (2) designing the action space, state space, and reward function specifically for unicast and broadcast scenarios in V2V communication. In \cite{b19}, a low-complexity multi-task dynamic offloading framework for S-MEC is introduced, dynamically allocating computing resources based on task numbers and dependencies. The authors in \cite{b20} have addressed the joint task offloading and resource allocation (JTORA) problem, optimizing task offloading, uplink power, and resource allocation at MEC servers.

While these works may have simulated mobile environments, none have proposed a mobility model for user equipment. In contrast, \cite{b21} introduces two key contributions: a TMSS-based mobile access prediction method leveraging user mobility patterns to enhance prediction accuracy, validated with human mobility data, and used for cloudlet reliability estimation in offloading decisions. Additionally, it proposes an integer encoding-based adaptive genetic algorithm for optimal offloading, considering cloudlet reliability, computation requirements, and mobile device energy consumption. The work in \cite{b22} tackles the challenge of large action spaces in resource allocation by combining reinforcement learning (RL) with heuristic information to speed up learning. The work in \cite{b23} addresses a mobile intelligent vehicle offloading scenario, where vehicles offload tasks to other vehicles or nearby infrastructure by exchanging resource information and making dynamic real-time decisions. The mobility model is based on location forecasting and uses the EKF algorithm to predict vehicle locations.

One common approach to modeling mobility is considering mobile equipment on a straight highway with no obstacles. This method often also models uniform acceleration for users. The works in \cite{b24, b25, b26} follow this model. The authors in \cite{b24} have proposed a method to measure the available computational resources of neighboring vehicles by calculating path connection and transmission times. The work in \cite{b25} addresses key challenges in vehicular task offloading, focusing on resource discovery, where varying vehicle mobility and computational capacities make finding reliable vehicles critical in dynamic, multi-hop networks. The authors in \cite{b26} have designed a game model to address competition and cooperation among the edge server, cloud, and requesting vehicle during task offloading and transformed the problem of maximizing the system's utility into a convex optimization problem. The work in \cite{b27} proposes a VEC-based computation offloading model that accounts for task data dependency and aims to minimize average response time and energy consumption, framed as a combinatorial optimization problem.

\subsection{Migration in Fog Computing Networks}

The authors in \cite{b28} have introduced Folo, a novel task allocation solution for vehicular fog computing (VFC), optimizing both latency and quality across stationary and mobile fog nodes. The task allocation process is formulated as a joint optimization problem, solved using LBO and a BPSO-based dynamic task allocation (DTA). The work in \cite{b29}, which is perhaps the most similar to ours, proposes a mobility-aware task offloading scheme for a two-layer fog system. It optimizes offloading and resource allocation based on user mobility to reduce task migration. A mobility model predicts UE sojourn times, while a utility function maximizes revenue by balancing energy consumption and delay. The problem is divided into two parts: a Gini coefficient-based offloading decision algorithm and a genetic algorithm for resource optimization. However, this model does not consider the cloud layer, which could be beneficial for tasks that require low energy consumption but are less sensitive to latency. Additionally, the sojourn time model may prove weak in real-world scenarios, where using a probabilistic model for mobility might not offer sufficient precision. The summary of related work is presented in \hyperref[table:related_works]{Table I}, and the main contributions of our study are demonstrated through a comparative analysis of various considerations. By employing a three-layered model, our approach offers enhanced flexibility in decision-making regarding offloading. Additionally, integrating a location and velocity trace model as part of the mobility model allows for the safe simulation of our proposed method in real-world scenarios. Considering migration further aids in optimal resource allocation and offloading.

\section{System Model}
\label{systemmodel}
In this section, we discuss the three-layered architecture and the model considered in this study. The architecture is composed of three primary components: user equipment (UE), fog nodes, and the cloud. A large number of UEs exist in the user layer, and they periodically release tasks that require computational resources. They are also mobile, and at different time steps, their positions and proximate fog nodes differ from one another. The fog layer consists of stationary fog computing nodes that provide much higher computing capacity at the cost of transmission delay and energy consumption. Every fog node has a coverage area in which it can receive requests from UEs. Unfortunately, due to the mobility of the UEs, we also have to consider the extra cost added to the system from migration, in which a UE might leave the coverage area of the fog node it has offloaded to. Additionally, the cloud layer exists in a centralized manner, providing an infinite amount of computing capacity at the cost of high propagation delay. It is also assumed that the updating process of UE mobility attributes and migration is handled by the cloud. The mobility update packets and the tasks that need migration are delivered to the fog nodes via the cloud layer.

We categorize the network components into user nodes, fog nodes, and the centralized cloud. The set of user nodes is defined as \(\mathcal{N}^u = \{ N^u_{1}, N^u_{2}, \dots, N^u_{U} \}\), and the set of fog nodes is defined as \(\mathcal{N}^f = \{ N^f_{1}, N^f_{2}, \dots, N^f_{F} \}\). The set of tasks generated by a UE is defined as \(\mathcal{T}_{i} = \{\tau_{1i}, \tau_{2i}, \dots\}\), and each task \(\tau_{ki}\) is described by a tuple \(\{\epsilon_{ki}, D_{ki}, \lambda^T_{ki}\}\), where \(\epsilon_{ki}\) denotes the number of CPU cycles required to execute one bit, \(D_{ki}\) represents the data size of the task in bits, and \(\lambda^T_{ki}\) is the latency sensitivity parameter used to calculate cost, as explained in the following sections. Tasks are generated periodically based on a fixed period and aperiodically following a exponential distribution. The details of the computation models are explained in the subsequent subsections. The main notations used throughout the paper are summarized in \hyperref[table:notation]{Table II}.

\subsection{Local Computation Model}

  According to many previous models \cite{b16,b29}, $f_{ki}$ can be computed as $f_{ki} = \epsilon_{ki}D_{ki}$ which is the CPU cycles needed to execute $\tau_{ki}$. Based on this task model, computation latency and transmission latency can be calculated using $f_{ki}$ and $D_{ki}$, respectively. For local execution, we have:

\begin{equation}
    T^{local}_{ki} = \frac{f_{ki}}{c^u_{i}}, \;\; \forall {\tau}_{ki} \in \mathcal{T}_i
\label{eq:1}
\end{equation}

where $c^u_{i}$ is the total computation power of $N^u_{i}$. Local servers are assumed to employ a queue-based execution model to represent execution over a time span. Accordingly, the waiting time \(T^{\text{wait}}_{ki}\) denotes the duration $\tau_{ki}$ must remain in the queue before being allocated the required resources. It should be noted that the precise value of this delay can be computed based on the state information of each individual server. A common method to calculate the energy consumption of execution is using the formula $E = \kappa c^2$ \cite{b30,b31,b32}, where $\kappa$ is the effective switched capacitance, which depends on the chip architecture \cite{b33}, and $c$ is the frequency allocated for the execution of the task. This formula gives the energy consumed per cycle; therefore, the local energy consumption to execute the task is calculated as follows:

\begin{equation}
    E^{local}_{ki} = \kappa(c^u_{i})^2f_{ki}, \;\; \forall {\tau}_{ki} \in \mathcal{T}_i
\label{eq:2}
\end{equation}

Based on equations (\ref{eq:1}) and (\ref{eq:2}), and given that there is no need to transmit the task in the local computation model, the total cost of executing $\tau_{ki}$ locally can be described as follows:

\begin{equation}
    \phi^{local}_{ki} = \lambda^{T}_{ki}(T^{local}_{ki} + T^{wait}_{ki}) + (1-\lambda^{T}_{ki})E^{local}_{ki}
\label{eq:3}
\end{equation}
where $\lambda^{T}_{ki}$ is the constant weight coefficient that determines the sensitivity of the task to execution delay, and $\lambda^{T}_{ki} \in [0,1]$. If $\lambda^{T}_{ki} > 0.5$, the task will become more sensitive to execution delay, such as in video streaming or communication tasks. If $\lambda^{T}_{ki} < 0.5$, the task will become more sensitive to energy consumption. In extreme cases, the value of $\lambda^{T}_{ki}$ can also be $0$ or $1$, indicating that the task's cost ignores execution delay or energy consumption, respectively.

\subsection{Fog Computation Model}
Fog nodes accept tasks from UEs and allocate a certain amount of resources for task execution. The process of offloading a task to a fog node involves a UE selecting the fog node as the destination, transmitting the task data to the chosen fog node using a wireless link, and executing the task with the allocated resources. In the fog computing model, the transmission of results back to the UE is not considered due to the insignificance of the data size and the superior channel conditions of the downlink compared to the uplink. In this model, we assume that a fog node can connect with multiple UEs simultaneously as long as they are within the service coverage of the fog node, as OFDMA (orthogonal frequency-division multiple access) technology enables this \cite{b34}. However, each UE can only connect with one fog node at a time, even if it is within the overlap of multiple service coverages. To calculate the energy consumption and latency of the transmission between a user and fog node, we consider the signal-to-interference-plus-noise-ratio (SINR) between the $i$-th UE and the $j$-th fog node as \(SINR_{ij} = \frac{P^u_{i}}{\sigma^2 + I_{ij}}\) \cite{b35}, where \(P^u_{i}\) is the total transmission power of the \(i\)-th UE node, \(\sigma^2\) is the noise power, and \(I_{ij}\) is the interference caused by other UEs using the same channel as \(N^u_{i}\). The transmission rate from \(N^u_{i}\) to \(N^f_{j}\) can then be expressed as:

\begin{equation}
    r_{ij} = W \log_2(1 + SINR_{ij}), \;\; \forall {N}^u_{i} \in \mathcal{N}^u, \forall {N}^f_{j} \in \mathcal{N}^f
\label{eq:4}
\end{equation}
where $W$ is the bandwidth of the wireless channels. Given the transmission rate between \(N^u_{i}\) and \(N^f_{j}\), the latency for transmitting $\tau_{ki}$ from user node \(N^u_{i}\) to fog node \(N^f_{j}\) can be calculated as follows:
\begin{equation}
    T^{trans}_{kij} = \frac{D_{ki}}{r_{ij}}, \;\; \forall \tau_{ki} \in \mathcal{T}_i, \forall {N}^f_{j} \in \mathcal{N}^f
\label{eq:5}
\end{equation}
Furthermore, given the time and power allocated for transmission, the energy consumed during transmission can be calculated as follows:
\begin{equation}
    E^{trans}_{kij} = P^u_iT^{trans}_{kij}, \;\; \forall \tau_{ki} \in \mathcal{T}_i, \forall {N}^f_{j} \in \mathcal{N}^f
\label{eq:6}
\end{equation}
Each task can be offloaded to either the fog or the cloud upon release. This decision, known as the offloading decision, is represented for $\tau_{ki}$ by $s_{ki} \in \{0, 1\}$. Furthermore, the set of all offloading decisions can be expressed as follows:
\begin{equation}
    S = \{ s_{ki} | \;\; \tau_{ki} \in \mathcal{T}_i, \;\; {N}^u_{i} \in \mathcal{N}^u \}
\label{eq:7}
\end{equation}
In the context of offloading destinations, we consider two possible options: a fog node or the cloud. For offloading to a fog node, the selection process is represented by $a_{kij} \in \{0, 1\}$, where $a_{kij}$ denotes the selection of the $j$-th fog node by $\tau_{ki}$. Hence, we can define the set of all fog node selections as follows:
\begin{equation}
    A^f = \{ a_{kij} | \;\; \tau_{ki} \in \mathcal{T}_i,\;  {N}^u_{i} \in \mathcal{N}^u, \;  {N}^f_{j} \in \mathcal{N}^f \}
\label{eq:8}
\end{equation}

After making the offloading decision, the algorithm allocates a certain amount of resources for the offloaded task. This resource allocation occurs only when the offloading destination is a fog node. The set of allocations for the $j$-th fog node is as follows:

\begin{equation}
    C_{j} = \{ c^f_{kij}|\;\; \tau_{ki} \in \mathcal{T}_i, \; {N}^u_{i} \in \mathcal{N}^u, \; {N}^f_{j} \in \mathcal{N}^f \}
    \label{eq:9}
\end{equation}

Each fog node is characterized by a maximum computing resource capacity, denoted by $c^f_j$, which varies across different nodes. The resource allocation must ensure that it stays within this limit. This is expressed by the following constraint:

\begin{equation}
    c^f_{kij} \leq c^f_{j}
\label{eq:10}
\end{equation}
Additionally, the complete set of allocation strategies can be expressed as:
\begin{equation}
    C = \bigcup^{F}_{j=1} C_{j}
\label{eq:11}
\end{equation}
Similar to the local execution model discussed in the previous subsection, the computation time for execution can be written as follows:
\begin{equation}
    T^j_{ki} = \frac{f_{ki}}{c^f_{kij}}, \;\; \forall \tau_{ki} \in \mathcal{T}_i, \forall {N}^f_{j} \in \mathcal{N}^f
\label{eq:12}
\end{equation}

The wait time for the task is also calculated in a similar way and is described as $T^{wait}_{kij}$. The energy consumption of executing the task on the $j$-th fog node is written as:
\begin{equation}
    E^{j}_{ki} = \kappa(c^f_{kij})^2f_{ki}, \;\; \forall \tau_{ki} \in \mathcal{T}_i, \forall {N}^f_{j} \in \mathcal{N}^f
\label{eq:13}
\end{equation}
Finally, the total cost of $\tau_{ki}$ when offloaded to the $j$-th fog node can be computed as follows:
\begin{equation}
\phi^j_{ki} = \lambda^T_{ki}(T^j_{ki}\!+\!T^{\text{trans}}_{kij}\!+\!T^{\text{wait}}_{kij}) + (1\!-\!\lambda^T_{ki})(E^{\text{trans}}_{ij}\!+\!E^j_{ki})
\label{eq:14}
\end{equation}

\begin{table}[t]
    \centering
    \captionsetup{justification=centering, labelsep=newline}
    \caption{Main Notation Used Throughout The Paper}
    \begin{tabular}{>{\centering\arraybackslash}m{2cm}>{\arraybackslash}m{6cm}}
        
        \hline
        \textbf{Symbol} & \textbf{Meaning} \\ 
        \specialrule{0.75pt}{0pt}{0pt} 
        \hline
        $U$ & Number of UEs \\ \hline
        $F$ & Number of fog nodes \\ \hline
        $N^u_{i}$ & $i$-th user node \\ \hline
        $N^f_{j}$ & $j$-th fog node \\ \hline
        $\mathcal{N}^{u}$ & Set of UEs, \(\mathcal{N}^u = \{ N^u_{1}, N^u_{2}, \dots N^u_{U} \}\)  \\ \hline
        $\mathcal{N}^{f}$ & Set of fog nodes, \(\mathcal{N}^f = \{ N^f_{1}, N^f_{2}, \dots N^f_{F} \}\)\\ \hline
        $\mathcal{N}^{uf}_i$ & Set of candidate fog nodes of $N^{u}_i$\\ \hline
        $\mathcal{N}^{fu}_j$ & Set of approximate UEs of $N^{f}_j$ \\ \hline
        $\tau_{ki}$ & The $k$-th task of $i$-th UE  \\ \hline
        $\mathcal{T}_i$ & The task set of $i$-th UE, $\mathcal{T}_{i} = \{\tau_{1i}, \tau_{2i}, \dots\}$  \\ \hline
        $D_{ki}$ & The data size of $\tau_{ki}$  \\ \hline
        $\epsilon_{ki}$ & The computation workload of $\tau_{ki}$\\\hline
        $\lambda^T_{ki}$ & The latency coefficient of $\tau_{ki}$ \\ \hline
        $f_{ki}$ & The CPU cycles required to execute $\tau_{ki}$  \\ \hline
        $\kappa$ & Effective switched capacitance \\ \hline
        $\sigma^2$ & The background noise variance \\ \hline
        $I_{ij}$ & The interference caused by UEs using the same channel \\ \hline
        $P^u_i$ & Total transmission power of user $N^u_i$ \\ \hline
        $W$ & Total bandwidth of transmission channel \\ \hline
        $S$ & Set of all offloading decisions \\ \hline
        $A^f$ & Set of all fog node selections \\ \hline
        $C_{j}$ & Set of allocations for the fog node $N^f_j$ \\ \hline
        $C$ & The complete set of allocation strategies \\\hline
        $A^c$ & Set of all cloud selections \\  \hline
        $r_{fc}$ & Transmission rate of wired links between fog nodes and the cloud \\ \hline
        $\delta$ & Migration cost coefficient \\  \hline                   
    \end{tabular}
    \label{table:notation}
    \vspace{-1.75em}
\end{table}

\subsection{Cloud Computation Model}
Offloading tasks to the cloud layer is also considered in our system model, as fog computing is intended to complement rather than replace cloud computing. In scenarios where fog computing resources are insufficient or energy preservation is a concern, offloading tasks to the cloud can be advantageous. The cloud selection is represented by \(a_{kic} \in \{0,1\}\), where \(a_{kic}\) denotes the selection of the cloud by the \(k\)-th task of $i$-th UE. The set of all cloud selections can be defined as:

\begin{equation}
    A^c = \{ a_{kic} | \;\; \tau_{ki} \in \mathcal{T}_i, \; {N}^u_{i} \in \mathcal{N}^u \}
\label{eq:15}
\end{equation}

The process of offloading to the cloud consists of first offloading the task to the $j$-th fog node using a wireless channel, and then transferring it to the cloud layer via a wired link. The cloud center is assumed to be located at a greater distance from the user equipment (UEs) compared to all fog nodes. Consequently, offloading tasks to the cloud results in higher transmission latency than selecting any fog node. This increased latency is primarily due to the propagation delay in the backbone network, which significantly contributes to the longer response times associated with cloud computing relative to fog computing. However, the cloud center typically offers more powerful computing capabilities compared to fog nodes, rendering the execution time negligible. Therefore, a task is considered complete once it is successfully transmitted to the cloud. We denote the transmission rate of the fog-to-cloud wired link by \(r_{fc}\) \cite{b15}; therefore, the delay of offloading to the cloud will be computed as follows:

\begin{equation}
    T^{trans}_{kic} = T^{trans}_{kij} + \frac{D_{ki}}{r_{fc}}, \;\; \forall \tau_{ki} \in \mathcal{T}_i, \forall {N}^f_{j} \in \mathcal{N}^f
\label{eq:16}
\end{equation}

Therefore, the total cost of offloading and executing $\tau_{ki}$ in the cloud can be calculated as follows:

\begin{equation}
    \phi^{c}_{ki} = \lambda^{T}_{ki}T^{trans}_{kic} + (1-\lambda^{T}_{ki})E^{trans}_{kij}
\label{eq:17}
\end{equation}

\subsection{Mobility Model}
\label{sec:mobility_model}
The movement area of the mobile user equipment (UE) is modeled as a rectangular region containing designated roads and paths along which each UE can traverse. At any time $t$, the mobility model of ${N}^u_{i}$ is represented by a 4-tuple $\{x^u_i(t), y^u_i(t), v^u_i(t), \alpha^u_i(t)\}$.
Here, $x^u_i$ and $y^u_i$ denote the UE's current position on the x-axis and y-axis, respectively, $v^u_i$ is the UE's speed, and $\alpha^u_i$ is the direction of movement in radians. The stationary fog nodes are represented as a 3-tuple $\{x^f_j, y^f_j, \rho^f_j\}$, where $x^f_j$ and $y^f_j$ are the fixed coordinates of the fog node, and $\rho^f_j$ is its coverage radius.

\subsection{Migration Model}
Given the mobility model, the migration model determines whether the UE remains within the coverage area of a fog node for the duration of the task. Migration occurs only if \( s_i = 1 \) and \( a_{ij} = 1 \), indicating that the task has been offloaded to the fog layer. To model this, we introduce the concept of sojourn time, defined as the duration for which the \( i \)-th UE stays inside the coverage area of the \( j \)-th fog node. Using the sojourn time \( T^{\text{soj}}_{ij} \), the migration variable for a task started at \( t_1 \) and expected to complete at \( t_2 \) is defined as:
\begin{equation}
    M_{kij}(t_1, t_2) =
    \begin{cases}
        0, & \text{if } T^{\text{soj}}_{ij} \geq (t_2 - t_1) \\
        1, & \text{if } T^{\text{soj}}_{ij} < (t_2 - t_1)
    \end{cases}
    \label{eq:18}
\end{equation}
The migration cost of \( \tau_{ki} \), as in previous works~\cite{b29}, can be modeled as a fraction of the data size. If \( \delta \) denotes the migration cost coefficient, then the migration cost is given by \( \phi^{\text{mig}}_{ki} = \delta D_{ki} \). Accordingly, the fog offloading cost is updated as:
\begin{equation}
\bar{\phi}^j_{ki} = \phi^{j}_{ki} + \phi^{\text{mig}}_{ki} \cdot M_{kij}(t_1, t_2)
\label{eq:19}
\end{equation}

\section{Problem Formulation \& Solution}
\label{problemFormulationAndSolution}

In this section, the problem is formulated and then solved using MOFCO, a heuristic-aided evolutionary game theory algorithm which is explained in the following subsections.
\subsection{Problem Formulation}
We aim to solve a multi-objective optimization problem, which we have formulated using a weighted sum approach. The two primary objectives that need to be minimized are the total latency and total energy, which are expressed as costs in different offloading decision scenarios. According to equations (\ref{eq:3}), (\ref{eq:17}) and (\ref{eq:19}) we can write the cost of $\tau_{ki}$ as follows:

{\small
\begin{equation}
\Phi_{ki}(s_{ki}, a_{kij}, a_{kic}, c^f_{kij}) = \begin{cases}
\phi^{local}_{ki} &\!\!,~s_{ki} = 0 \\
\bar{\phi}^{j}_{ki} &\!\!,~s_{ki} = 1, a_{kij} = 1 \\
\phi^{cloud}_{ki} &\!\!,~s_{ki} = 1, a_{kic} = 1
\end{cases}
\label{eq:20}
\end{equation}
}

Given the definition of the overall cost function, the multi-objective optimization problem can be expressed as follows:

\begin{equation}
\begin{aligned}
\text{P1:} \quad &\min_{S, A^f, A^c, C} \sum_{N^u_i \in \mathcal{N}^u} \sum_{\tau_{ki} \in \mathcal{T}_i} \Phi_{ki}(s_{ki}, a_{kij}, a_{kic}, c^f_{kij}) \\
\text{s.t.} \quad &\text{C1:} \quad s_{ki} = \{0, 1\}, \quad \forall \tau_{ki} \in \mathcal{T}_i \\
&\text{C2:} \quad a_{kij} = \{0, 1\}, \quad \forall \tau_{ki} \in \mathcal{T}_i, \;\; \forall N^f_j \in \mathcal{N}^f \\
&\text{C3:} \quad a_{kic} = \{0, 1\}, \quad \forall \tau_{ki} \in \mathcal{T}_i \\
&\text{C4:} \; \sum^{F}_{j=1} a_{kij} + a_{kic} \leq 1, \; \forall \tau_{ki} \in \mathcal{T}_i, \; \forall N^f_j \in \mathcal{N}^f\\
&\text{C5:} \quad c^f_{kij} \leq c^f_j, \quad \forall \tau_{ki} \in \mathcal{T}_i, \;\; \forall N^f_j \in \mathcal{N}^f
\end{aligned}
\label{eq:21}
\end{equation}

The first three constraints which are C1, C2, and C3 indicate the binary values of the offloading decision, choosing the fog layer and the cloud layer respectively. Constraint C4 indicates that every UE can only offload to one single destination or execute the task locally. Constraint C5 indicates that allocated resource of tasks offloaded to a single fog node does not exceed the maximum capacity of that fog node.

This problem is a mixed-integer nonlinear programming problem that is NP-hard and to solve it we have conducted a heuristic-aided evolutionary game theory approach called MOFCO, that tries to eliminate as many fog nodes as it can regarding the lowest cost that each fog node can provide and then, it explores the remaining solution space finding the solution with fewer migrations and reduces the final cost of each task. MOFCO utilizes mobility and workload prediction to reduce the overall cost of migration and execution. The following subsections present the mobility and workload prediction models, followed by a detailed explanation of MOFCO.

\begin{figure}
\centerline{\includegraphics[width=\columnwidth]{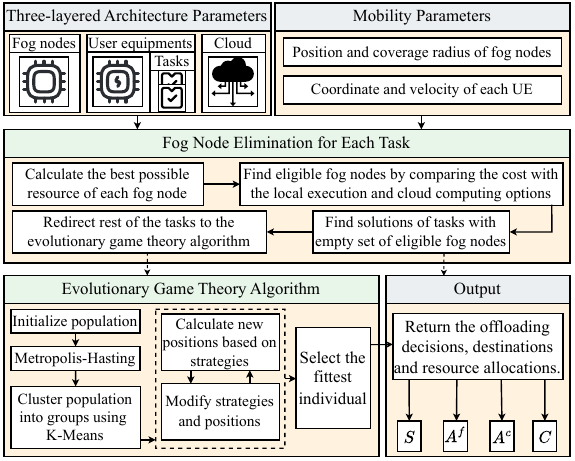}}
\caption{Overall structure of our proposed method, MOFCO}
\label{proposedFig}
\vspace{-1.75em}
\end{figure}

\subsection{Mobility \& Workload Prediction}
Based on the mobility model discussed in Section~\ref{sec:mobility_model}, we can find the time that $N^u_i$ exits the coverage of $N^f_j$ by solving a quadratic equation. Let the initial UE position at \(t_1\) be \((x^u_i(t_1), y^u_i(t_1))\), the direction be \(\alpha^u_i(t_1)\), and the constant speed be \(v^u_i(t_1)\). The position of the UE at any time \(t = t_1 + \varsigma\) can be predicted as:
\begin{equation}
    x^u_i(t_1 + \varsigma) = x^u_i(t_1) + v^u_i(t_1) \cos(\alpha^u_i(t_1)) \cdot \varsigma
    \label{eq:22}
\end{equation}
\begin{equation}
    y^u_i(t_1 + \varsigma) = y^u_i(t_1) + v^u_i(t_1) \sin(\alpha^u_i(t_1)) \cdot \varsigma
    \label{eq:23}
\end{equation}

To find the time \(\varsigma_{s}\) at which the UE exits the coverage of fog node \(j\), we solve the following equation:
\begin{equation}
    (x^u_i(t_1 + \varsigma) - x^f_j)^2 + (y^u_i(t_1 + \varsigma) - y^f_j)^2 = (\rho^f_j)^2
    \label{eq:24}
\end{equation}

Substituting from (\ref{eq:22}) and (\ref{eq:23}), we get:
\begin{align}
    &[x^u_i(t_1) + v^u_i \cos(\alpha^u_i) \cdot \varsigma - x^f_j]^2 + \nonumber \\
    &[y^u_i(t_1) + v^u_i \sin(\alpha^u_i) \cdot \varsigma - y^f_j]^2 = (\rho^f_j)^2
    \label{eq:25}
\end{align}

Let us define:
\[
\Delta x = x^u_i(t_1) - x^f_j,\quad \Delta y = y^u_i(t_1) - y^f_j
\]

This yields the quadratic equation in \(\varsigma\):
\begin{align}
   & (v^u_i)^2 \varsigma^2 + 2v^u_i (\Delta x \cos\alpha^u_i + \Delta y \sin\alpha^u_i)\varsigma +  \nonumber \\
   & (\Delta x)^2 + (\Delta y)^2 - (\rho^f_j)^2 = 0
   \label{eq:26}
\end{align}

The sojourn time \(\varsigma^{ij}_s\) is the smallest non-negative root of this equation:
\begin{align}
    &\varsigma^{ij}_s = \frac{-B + \sqrt{B^2 - 4AC}}{2A}, \quad \text{where: } \\
    &A = (v^u_i)^2, \nonumber \\ 
    &B = 2v^u_i (\Delta x \cos\alpha^u_i + \Delta y \sin\alpha^u_i), \nonumber \\
    &C = (\Delta x)^2 + (\Delta y)^2 - (\rho^f_j)^2 \nonumber
    \label{eq:27}
\end{align}

We select the positive root because we are only interested in the future time when the UE leaves the coverage area of the fog node. Based on previous models such as \cite{b29} the sojourn can be modeled using an exponential distribution so we can now find the migration probability according to equation (\ref{eq:18}):
\begin{equation}
    \mathbb{P}(M_{ij}^k(t_1, t_2) = 1) =  1 - e^{-\dfrac{t_2}{\varsigma^{ij}_s}}
    \label{eq:28}
\end{equation}

This represents the probability that $N^u_i$ will have exited $N^f_j$’s coverage area by time \(t_2\), triggering a migration event. One of the main contributions of this article is the prediction of workload based on the current sojourn times of the UEs. To estimate the workload of a fog node after $t$ seconds, we utilize the available sojourn time information using the following equation:

\begin{equation}
    W_j(t) = \min\left(\Theta^j_s,\; t \cdot |\mathcal{N}^{uf}_j|\right)
    \label{eq:29}
\end{equation}

Here, $\mathcal{N}^{uf}_j$ denotes the set of UEs currently within the coverage radius of fog node $N^f_j$, and $\Theta^j_s$ is the total sojourn time of UEs associated with $N^f_j$, computed as:

\begin{equation}
    \Theta^j_s = \sum_{N^u_i \in \mathcal{N}^{fu}_j} \varsigma^{ij}_s
    \label{eq:30}
\end{equation}

This parameter is used to evaluate potential destination fog nodes and select the most suitable one based on the current state of the system. This approach improves fog node utilization and enhances load balancing across the network.

\subsection{Proposed Method}
In this section, we describe the proposed method to solve this multi-objective MINLP problem. The primary approach is a two-step heuristic and evolutionary game theory driven method illustrated in Fig. \ref{proposedFig} that limits the search space of the problem by calculating the best case of each proximate fog node and eliminating them from offloading options, and then the remaining solution space is searched using the evolutionary game theory approach and by leveraging randomized search patterns, it can efficiently choose fog nodes as offloading destination and determine the allocated resource of the task. Algorithm (\ref{alg:1}) illustrates our proposed method which is run for every task $\mathcal{T}_t$ released at timestep $t$.

\begin{algorithm}[t]
\fontsize{8}{8}\selectfont
\caption{The MOFCO algorithm}
\textbf{Input:} $\mathcal{T}_t, p, g, k, f_{eval}, l_{thresh}$ \\
\textbf{Output:} $S, A^f, A^c, C$
\begin{algorithmic}[1]
\State \textbf{Step 1 Fog node elimination:}
\State Initialize: $\mathcal{T} = \emptyset$ // Set of unassigned tasks
\For{$\tau_{ki}$ in $\mathcal{T}_t$}:
    \State Initialize: $\mathcal{N}^{efu}_i = \emptyset$; // Set of eligible fog nodes
    \For{$N^f_j$ in $N^{fu}_i$}: // For each candidate fog node of the task
        \If{$\phi^j_{ki}(c^{f,\text{opt}}_{kij}) \leq min(\phi^c_{ki}, \phi^{local}_{ki})$ }: // Based on eq. (\ref{eq:34})
            \State $\mathcal{N}^{efu}_i$.add($N^f_j$);
        \EndIf
    \EndFor
    \If{$\mathcal{N}^{efu}_i == \emptyset$}:
        \If{$\phi^{local}_{ki} > \phi^{cloud}_{ki}$}:
            \State $s_{ki} = 1$;
            \State $a^c_{ki} = 1$;
        \Else:
            \State $s_{ki} = 0$;
        \EndIf
    \Else
        \State $\mathcal{T}.add(\tau_{ki})$;
    \EndIf
\EndFor
\State \textbf{Step 2 Using evolutionary game theory to find solution for remaining UEs:}
\If{$\mathcal{T} \neq \emptyset$}
    \State EvolutionaryGameTheory($\mathcal{T}$, $p$, $g$, $k$, $f_{eval}$, $l_{thresh}$);
\EndIf
\end{algorithmic}
\label{alg:1}
\end{algorithm}

To select eligible fog nodes, we estimate the minimum cost each node can offer by ignoring migration and optimizing the allocated resource \( c^f_{kij} \) using Equation~\eqref{eq:14}. The total cost of offloading task \( \tau_{ki} \) to fog node \( N^f_j \) is:

\begin{align}
    &\phi^j_{ki} = \lambda^T_{ki}(\frac{f_{ki}}{c^f_{kij}} + T^{\text{trans}}_{kij} + T^{\text{wait}}_{kij}) + \nonumber \\
    & (1 - \lambda^T_{ki})\left(E^{\text{trans}}_{kij} + \kappa (c^f_{kij})^2 f_{ki} \right)
    \label{eq:31}
\end{align}

Only two terms that depend on $c^f_{kij}$ are the computation time and the fog execution energy. In order to simplify the process we isolate these and define:

\begin{equation}
\phi^j_{ki}(c^f_{kij}) = \lambda^T_{ki} \cdot \frac{f_{ki}}{c^f_{kij}} + (1 - \lambda^T_{ki}) \cdot \kappa (c^f_{kij})^2 f_{ki} + \text{const}
\label{eq:32}
\end{equation}

To find the optimal value of $c^{f,\text{opt}}_{kij}$ we take the derivative and set it to zero:

\begin{align}
\frac{d\phi^j_{ki}}{dc^f_{kij}} &= -\lambda^T_{ki} \cdot \frac{f_{ki}}{(c^f_{kij})^2} + 2(1 - \lambda^T_{ki}) \cdot \kappa f_{ki} \cdot c^f_{kij} = 0 \nonumber \\
\Rightarrow (c^f_{kij})^3 &= \frac{\lambda^T_{ki}}{2(1 - \lambda^T_{ki}) \kappa} \nonumber \\
\Rightarrow c^{f,\text{opt}}_{kij} &= \left( \frac{\lambda^T_{ki}}{2(1 - \lambda^T_{ki}) \kappa} \right)^{\frac{1}{3}}
\label{eq:33}
\end{align}

Now, by comparing the best possible cost of the fog node with the local and cloud costs, we can eliminate fog nodes if one of the following inequalities holds:

\begin{equation}
    \phi^j_{ki}(c^{f,\text{opt}}_{kij}) > \phi^c_{ki} , \quad \phi^j_{ki}(c^{f,\text{opt}}_{kij}) > \phi^{\text{local}}_{ki}
    \label{eq:34}
\end{equation}

This part of the algorithm ensures that fog nodes with high queue wait times are eliminated due to the high latency they impose on the tasks. Additionally, in some extreme \( \lambda^T_{ki} \) values, such as \( \lambda^T_{ki} = 0 \), fog nodes may be completely eliminated because of the superior energy efficiency of cloud computing. Lines 1-20 of algorithm (\ref{alg:1}) explain this process.

For the remaining tasks, we adopt a novel evolutionary game theory method based on the approach proposed in~\cite{b34}, which we adapt and optimize for our fog computing model. Traditional evolutionary algorithms often overlook the interactions among individuals; the changes in individuals to achieve better fitness are typically random and do not account for the influence of others or the current state of the population. To address this limitation, we incorporate structured populations and introduce a competitive mechanism in which each individual competes with the best member of its group to improve its fitness. This significantly enhances the performance of the algorithm.

Initially, as shown in line 1 of Algorithm~\ref{alg:2}, each individual is assigned a random fog node and resource allocation, along with a randomly initialized strategy vector for future steps. In line 2, each individual undergoes the Metropolis-Hastings algorithm to determine its fog node destination. To reduce the randomness in the initial population, this technique guides individuals to calculate the migration probability and the fog node evaluation parameter of candidate fog nodes prior to the main evolutionary iterations. In this step, the position (i.e., fog node assignment) of the individual is changed to a new random value. The fitness of both the original and new positions is evaluated, and the new position is accepted with the following probability:

\begin{equation}
    \min\left(\frac{f(Y')}{f(Y)}, 1\right),
    \label{eq:35}
\end{equation}

where $f$ is the fitness function, and $Y'$ denotes the new version of individual $Y$. Through this process, fog node destinations are refined. Subsequently, the K-means clustering algorithm is applied to group individuals into $k$ clusters (a predefined hyperparameter), based on Euclidean distance. This results in clusters of individuals with similar fog node destinations, which are then used for the competition phase.

In lines 4–7 of Algorithm~\ref{alg:2}, each individual competes to improve its fitness using a game-theoretic approach. The strategy of an individual $Y$ is denoted by $\sigma^Y$, a vector of values uniformly drawn from the interval $[0,1]$, one per task. In line 5, each individual updates its position based on its strategy as follows:

\begin{equation}
    pos^y_i = pos^y_i + R \cdot N(0, \sigma^y_i), \quad \forall \sigma^y_i \in \sigma^Y,
    \label{eq:36}
\end{equation}

where $position^y_i$ is the allocated resource for task $i$, $R$ is a scaling factor controlling the step size, and $N(0, \sigma^y_i)$ denotes a random value drawn from a normal distribution with mean 0 and standard deviation $\sigma^y_i$.

In line 6, a subset of individuals is selected, and their fitness is compared with the best individual in their respective clusters, denoted $Y^{\text{best}}$. If an individual is selected for modification, its strategy is replaced by that of $Y^{\text{best}}$, i.e., $\sigma^Y = \sigma^{Y^{\text{best}}}$, and its position is updated according to:

\begin{equation}
    pos^Y = pos^Y + r \cdot (pos^{Y^{\text{best}}} - pos^Y),
    \label{eq:37}
\end{equation}

where $r \sim \mathcal{U}[0.5, 1]$. This equation shifts the individual’s position (i.e., resource allocation) towards the cluster’s best solution by a random factor between 0.5 and 1.

Our approach achieves a balance between exploration—via randomized movement determined by individual strategies—and exploitation—by guiding individuals toward superior solutions within their clusters. This cluster-based exploitation enhances search efficiency by leveraging similarities among individuals.

To compute the reward for a task $\tau_{ki}$, we define the Fog Node Evaluation Factor (FNEF), which helps assess fog nodes based on expected future workload. This factor is calculated using the following heuristic rule derived from Eq.~(\ref{eq:30}):

\begin{equation}
    \text{FNEF}_j(f_{\text{eval}}, l_{\text{thresh}}, l) =
    \begin{cases}
        -\Theta^j_s \cdot f_{\text{eval}}, & l < l_{\text{thresh}} \\
        \phantom{-}\Theta^j_s \cdot f_{\text{eval}}, & l \geq l_{\text{thresh}}
    \end{cases}
    \label{eq:38}
\end{equation}

in which $f_{eval}$ is the coefficient that determines how much the workload prediction affects the reward, and $l_{thresh}$ is a latency threshold that distinguishes between low-latency and high-latency tasks. The FNEF reflects the following intuition:
\begin{itemize}
    \item Fog nodes with higher projected workloads are better suited for low-latency tasks.
    \item Fog nodes with lighter future workloads are preferred for high-latency tasks.
\end{itemize}

These heuristics promote efficient fog node utilization by assigning tasks based on workload characteristics, thus improving overall cost-effectiveness. Finally, the fitness (reward) of task $\tau_{ki}$—released at timestep $t$, assigned to fog node $N^f_j$ with expected latency $l$—is calculated as:

\begin{align}
    & f(\tau_{ki}, t, l_{\text{thresh}}, l, f_{\text{eval}}) = -\big( \phi^j_{ki} +\nonumber \\
    & \text{FNEF}(f_{\text{eval}}, l_{\text{thresh}}, l) + \nonumber \\
    & \mathbb{P}(M_{ij}^k(t, t+l) = 1) \cdot \delta \ \cdot D_{ki}\big) 
    \label{eq:39}
\end{align}

where $\phi^j_{ki}$ is the cost, and $\mathbb{P}(M_{ij}^k(t, t+l) = 1)$ denotes the probability that task $\tau_{ki}$ is migrated during the interval $[t, t+l]$, incurring a penalty $\delta \cdot D_{ki}$.

\begin{algorithm}[t]
\fontsize{8}{8}\selectfont
\caption{Evolutionary Game Theory Optimization Method}
\textbf{Input:} $\mathcal{T}$,  $p$, $g$, $k$, $f_{\text{eval}}$, $l_{\text{thresh}}$ \\
\textbf{Output:}  $A^f$, $C$
\begin{algorithmic}[1]
\State $\text{Population} \gets \textsc{InitializePopulation}(\mathcal{T}, p)$
\State $\text{Population} \gets \textsc{MetropolisHastings}(\text{Population}, f_{\text{eval}}, l_{\text{thresh}})$
\State $\text{Clusters} \gets \textsc{KMeans}(\text{Population}, k)$
\For{$i = 1$ to $g$}
    \State \textsc{UpdatePositions}$(\text{Clusters})$
    \State \textsc{ModifyStrategiesAndPositions}$(\text{Clusters}, f_{\text{eval}}, l_{\text{thresh}})$
\EndFor
\State $\text{BestIndividual} \gets \textsc{FindFittest}(\text{Clusters}, f_{\text{eval}}, l_{\text{thresh}})$
\State $(A^f, C) \gets \textsc{AssignDestinationAndResources}(\text{BestIndividual})$
\end{algorithmic}
\label{alg:2}
\end{algorithm}

\subsection{Time Complexity}
The worst-case scenario for the size of $\mathcal{T}_t$ is the number of UEs, denoted as $U$. The number of fog nodes is denoted by $F$. Hence, the fog node elimination process results in a complexity of $O(U \cdot F)$. In the evolutionary algorithm, considering the population size as $p$ and the number of steps as $g$, the population initialization takes $O(p \cdot U)$. The Metropolis-Hastings step performs constant-time operations per individual, resulting in $O(p \cdot U)$. The K-means clustering algorithm runs for a limited number of iterations, contributing $O(p \cdot k)$. The main loop executes $g$ times; thus, updating positions and modifying strategies requires $O(g \cdot p \cdot U)$. Finding the fittest individual takes $O(p \cdot U)$, and assigning the solution to each task takes $O(U)$. The total complexity is $O(U \cdot F) + O(p \cdot U) + O(p \cdot U) + O(p \cdot k) + O(g \cdot p \cdot U) + O(p \cdot U) + O(U)$. After eliminating the lower-order terms, the dominant terms are $\boldsymbol{O(U \cdot F) + O(g \cdot p \cdot U)}$.

\section{Simulation Results}
\label{simulationResults}
\begin{figure}[t]
\centerline{\includegraphics[width=\columnwidth]{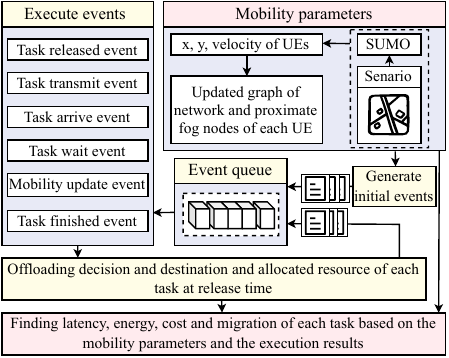}}
\caption{Overall structure of the event-based simulation process.
}
\label{eventBasedSimulation}
\vspace{-1.5em}
\end{figure}

In this section, the MOFCO algorithm is evaluated using an event-based simulation, as illustrated in Fig. \ref{eventBasedSimulation}. In this process, mobility parameters and tasks are generated for each UE. Then, these tasks generate events that are executed and produce other types of events, which are inserted into the event queue; the simulation executes events and injects new events back into the queue until it is empty. Various network parameters were tested, and the results show that MOFCO reduces the overall cost.

\subsection{Evaluation Setup}
Our simulation was developed using Python 3.11.0, and the mobility model and physical environments were taken from the results of pre-built scenarios from SUMO. The physical environment of the simulation is based on the open-source scenario TAVF-Hamburg \cite{b36}. The output of the SUMO simulation includes the exact position and velocity of each vehicle at every time step (in seconds). The vehicles were monitored over a timespan of 30 minutes, equal to 1800 seconds, and the compared algorithms used these vehicles as UEs. Each user equipment (UE) is assumed to generate two types of tasks: periodic and aperiodic. Periodic tasks are generated with a period \( T_{periodic} \in [30, 40] \), while aperiodic tasks are generated according to an exponential distribution with rate \( \lambda_{aperiodic} \in [0.05, 0.2] \). Consequently, the number of aperiodic tasks within any interval follows a Poisson distribution, a standard model for aperiodic task arrivals in many systems.
Other parameters of the simulation are defined in TABLE \ref{table:simconfig}. The values are taken from previous works \cite{b29, b30, b31, b32, b33, b35}.
The proposed algorithm is compared with four baseline algorithms. The first one is a greedy approach where the UEs only offload the tasks to the cloud layer (OnlyCloud). The second approach (OnlyLocal) avoids offloading and executes tasks locally. The third algorithm, referred to as RA, randomly decides whether to offload a task or execute it locally. If offloading is selected, the destination fog node is also chosen at random. To evaluate the impact of mobility and migration, the fourth approach, GCGA, is based on the prior work presented in~\cite{b29}. This method combines a Gini coefficient-based heuristic with a genetic algorithm and is fully aware of both mobility and migration dynamics, thereby accurately capturing the characteristics of our fog computing model.

\begin{figure*}[t]
    \centering
    \begin{subfigure}{0.49\textwidth}
        \centering
        \includegraphics[width=\textwidth]{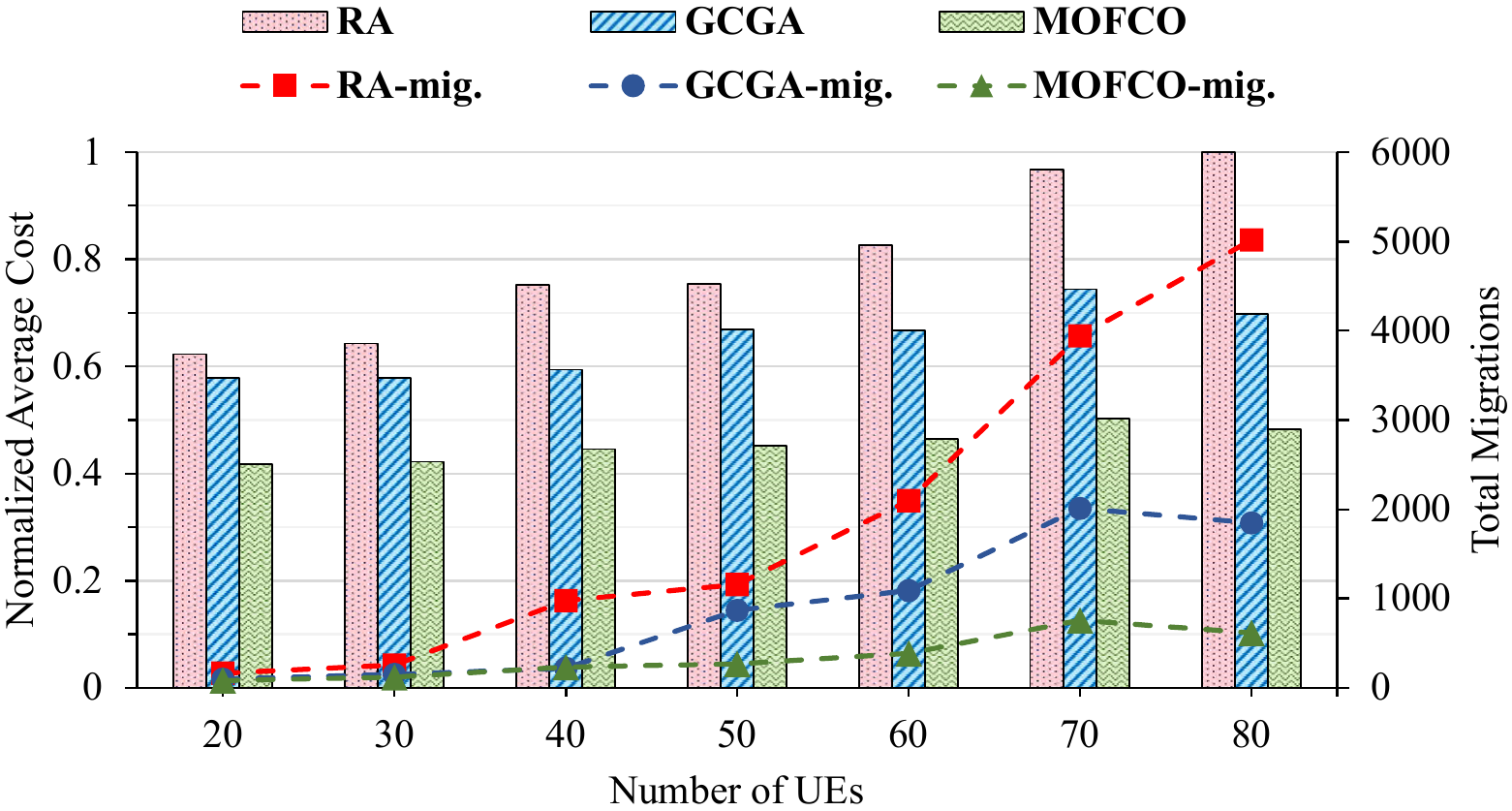} 
        \caption{}
    \end{subfigure}
    \hfill
    \begin{subfigure}{0.49\textwidth}
        \centering
        \includegraphics[width=\textwidth]{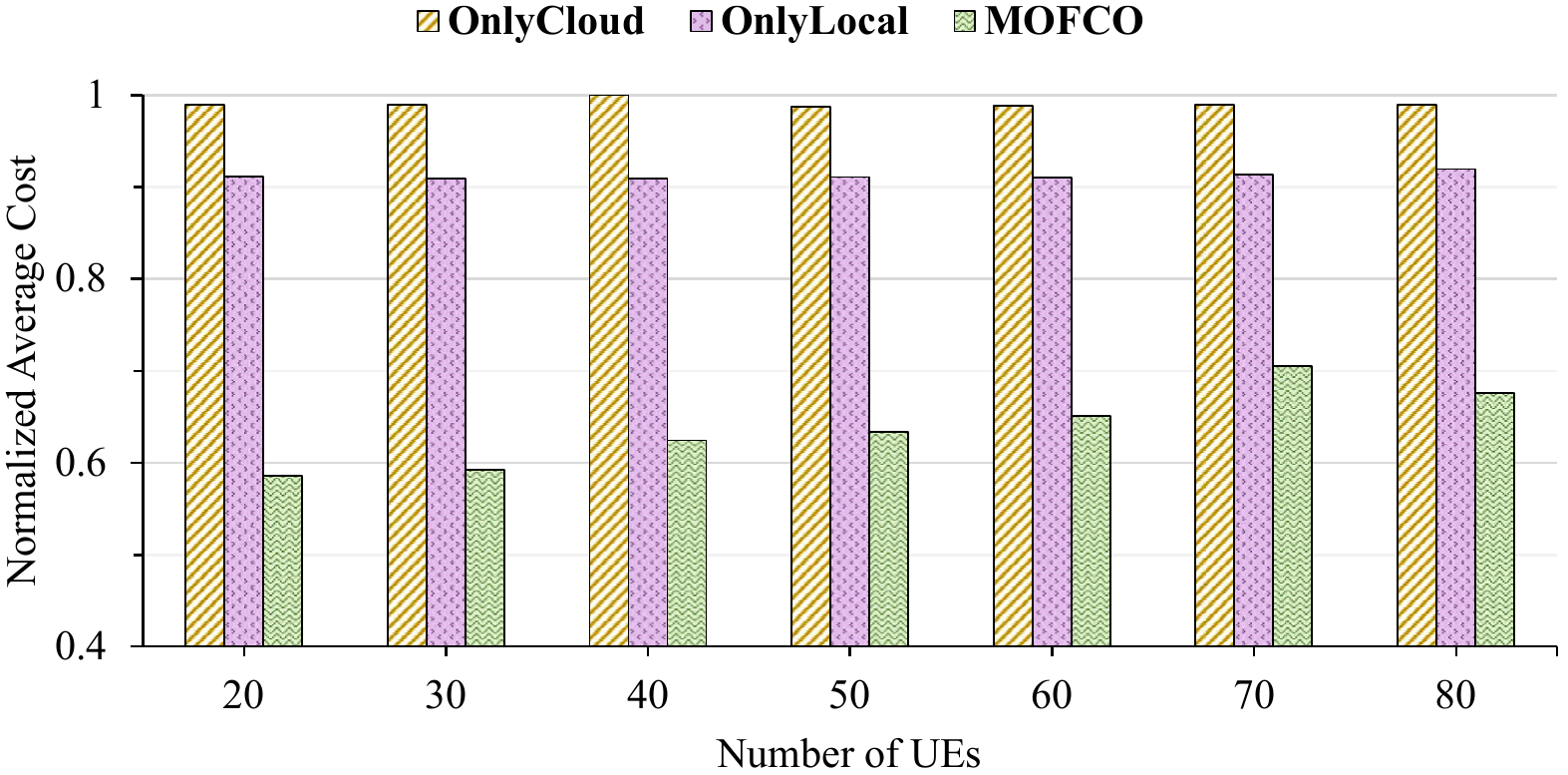} 
        \caption{}
    \end{subfigure}
    \caption{Simulation results over a timespan of 30 minutes with 35 fog nodes and default parameters. (a) Impact of UE count on total cost and migration count. (b) Impact of UE count on OnlyCloud and OnlyLocal.}
    \label{userCounts}
\vspace{-0.5em}
\end{figure*}

\begin{figure*}[t]
    \centering
    \begin{subfigure}{0.49\textwidth}
        \centering
        \includegraphics[width=\textwidth]{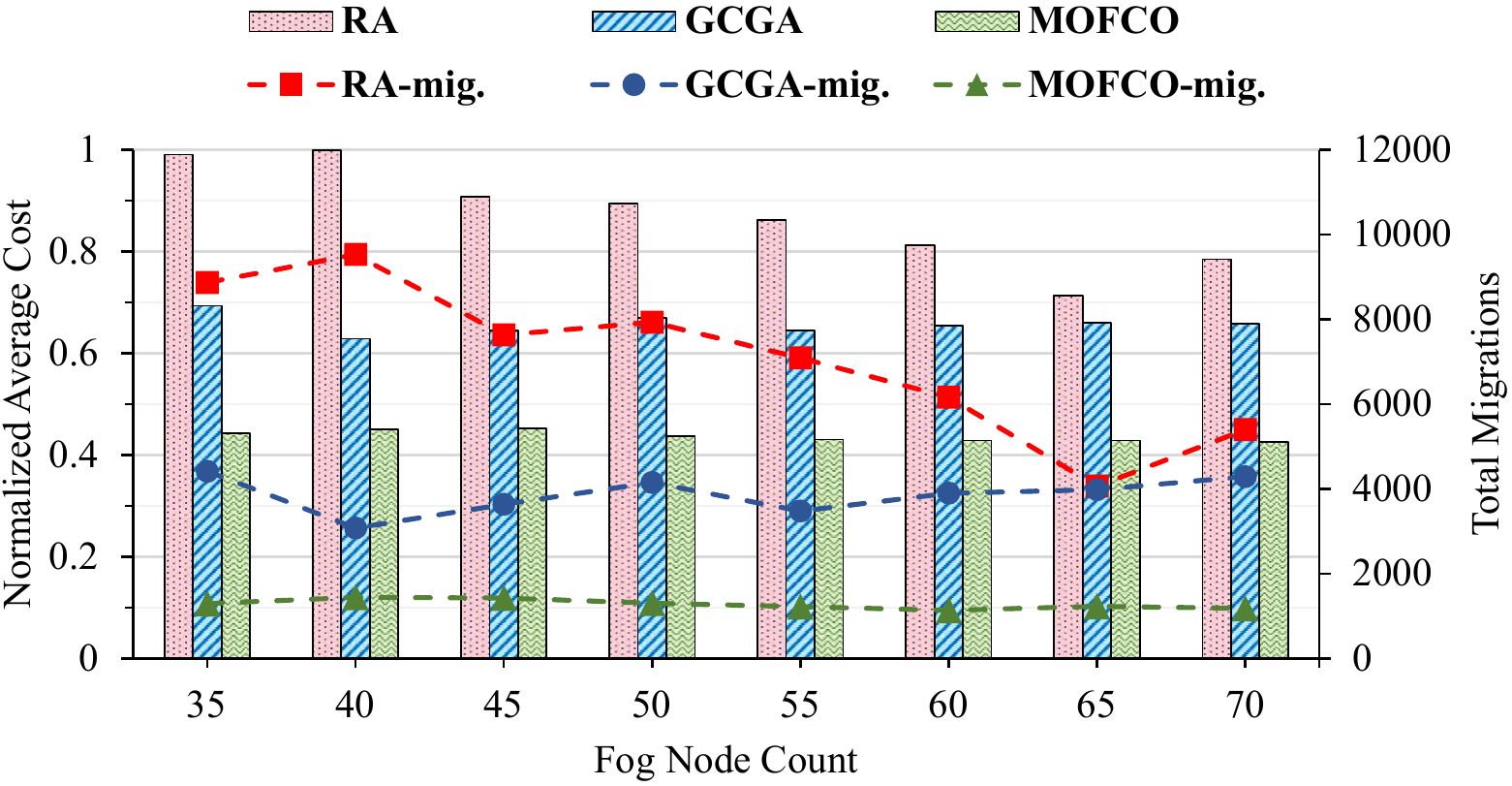} 
        \caption{}
    \end{subfigure}
    \hfill
    \begin{subfigure}{0.49\textwidth}
        \centering
        \includegraphics[width=\textwidth]{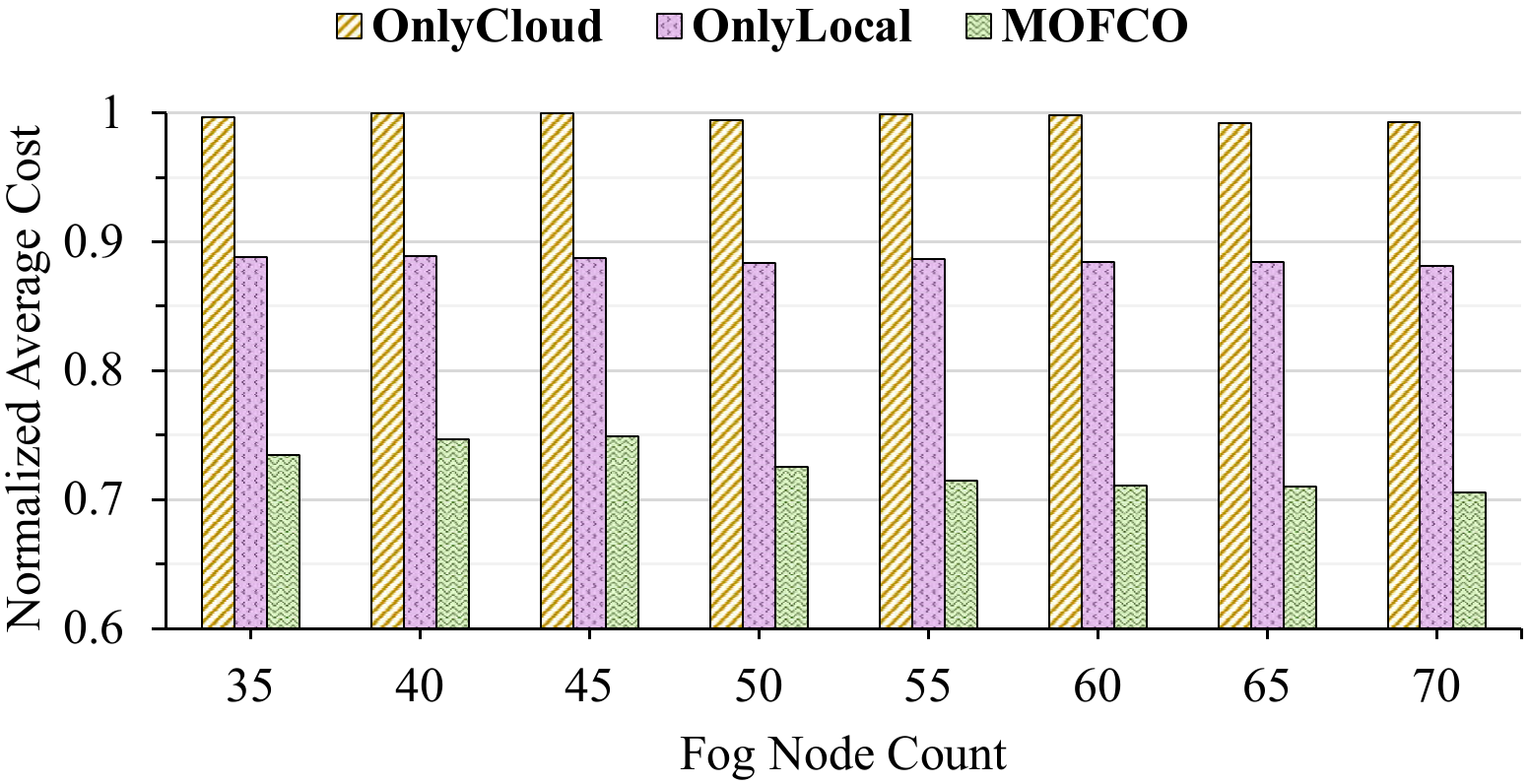} 
        \caption{}
    \end{subfigure}
    \caption{Simulation results over a timespan of 30 minutes with 70 UEs and default parameters. (a) Impact of fog node count on total cost and migration count. (b) Impact of fog node count on OnlyCloud and OnlyLocal.}
    \label{fogCounts}
\vspace{-1.25em}
\end{figure*}

\begin{table}[b]
\vspace{-1.75em}
\captionsetup{justification=centering, labelsep=newline}
\caption{{Simulation Configuration}}
\centering
\begin{tabular}{|p{1.5cm}p{2cm}|p{1.5cm}p{2cm}|}
\hline
\textbf{Parameter} & \textbf{Value} & \textbf{Parameter} & \textbf{Value} \\
\specialrule{0.75pt}{0pt}{0pt} 
\hline
$c^u_i$ & [2, 3] GHz 
&
$c^f_j$ & \scriptsize{[4, 6] GHz (default)} \\
 \hline
$\sigma^2$ & -174 dbM 
&
$I_{ij}$ & -75 dbM \\
\hline
$P^u_i$ & [80, 100] mW 
&
$W$ & [10, 20] MHz \\
\hline
$r_{fc}$ & 15 Mbps 
&
$D_i$ & [125, 175] Mb \\
\hline
$ T_{periodic}$ & [30, 40] second 
&
$\lambda_{aperiodic}$ & [0.05, 0.2] \\
\hline
$\delta$ & 1E-07 (default)
&
$\epsilon_i$ & \scriptsize{[30, 120] cycles/bit} \\
\hline
$U$ & 70 (default) 
&
$F$ & 35 (default)\\
\hline
$p$ &  40 
&
$g$ &  20 \\
\hline
$k$ & 5 
&
$\lambda^T$ & [0, 1]\\
\hline
$f_{eval}$ & 0.3 
&
$l_{thresh}$ & 4\\

\hline
\end{tabular}
\label{table:simconfig}

\end{table}

\subsection{Effect of Number of UEs}

To demonstrate the scalability of our method, we simulated a scenario with different numbers of UEs generating tasks over a time span of 30 minutes. In Fig.~\ref{userCounts}.a, the effect of the number of UEs on RA, GCGA, and MOFCO can be seen. The total number of migrations shows an increasing trend because the overall number of UEs increases. RA and GCGA have a high number of migrations, while MOFCO manages to keep the number of migrations lower, which eventually results in better cost compared to the other methods.
The normalized cost is calculated as the average cost per task, divided by the maximum task cost across all scenarios. MOFCO performs on average 19\% better than GCGA and achieves up to 24\% improvement in certain scenarios.
In Fig.~\ref{userCounts}.b, our method is compared with OnlyCloud and OnlyLocal. These two methods impose a similar total cost on the system, with OnlyLocal resulting in the highest energy consumption and OnlyCloud causing high latency. Our method performs better in all scenarios despite having non-zero migrations, because the other two do not fully utilize the system’s computational power. The trends for OnlyLocal and OnlyCloud remain unchanged as the number of UEs increases, since each task is deterministically computed on similar types of devices. MOFCO is compared with these two methods to show that our method fully utilizes the benefits of fog computing and decreases the downsides of the previous models, such as the high transmission delay of cloud computing.

\subsection{Effect of Fog Node Count}
In Fig.~\ref{fogCounts}.a, to further test scalability, we increased the total number of fog nodes in a scenario with 70 UEs and raised the aperiodic task generation rate to $\lambda_{aperiodic} \in [0.1, 0.3]$. We then recorded the costs of RA, GCGA, and MOFCO. As expected, with more fog nodes, the total computing capacity increases, resulting in a decreasing trend in overall cost. As shown in the figure, the total number of migrations in GCGA slightly decreases, keeping its cost lower than that of the RA algorithm in every scenario. However, MOFCO achieves the lowest cost among these algorithms and maintains the total number of migrations below 1300. Although the decreasing trend in migration is not clearly visible in the figure, the total number of migrations decreases from 1294 with 35 fog nodes to 1186 with 70 fog nodes. MOFCO performs on average 22\% better than GCGA and achieves up to 25\% improvement in certain scenarios.
In Fig.~\ref{fogCounts}.b, we compare MOFCO with OnlyCloud and OnlyLocal. Both of these algorithms result in higher costs and neither adjust task distribution across scenarios. As expected, their costs remain relatively constant. This comparison highlights that MOFCO effectively utilizes the benefits of fog computing, and the decreasing trend in cost is more visually apparent in this figure.

\subsection{Effect of Other Parameters}
\begin{figure}[t]
\centerline{\includegraphics[width=\columnwidth]{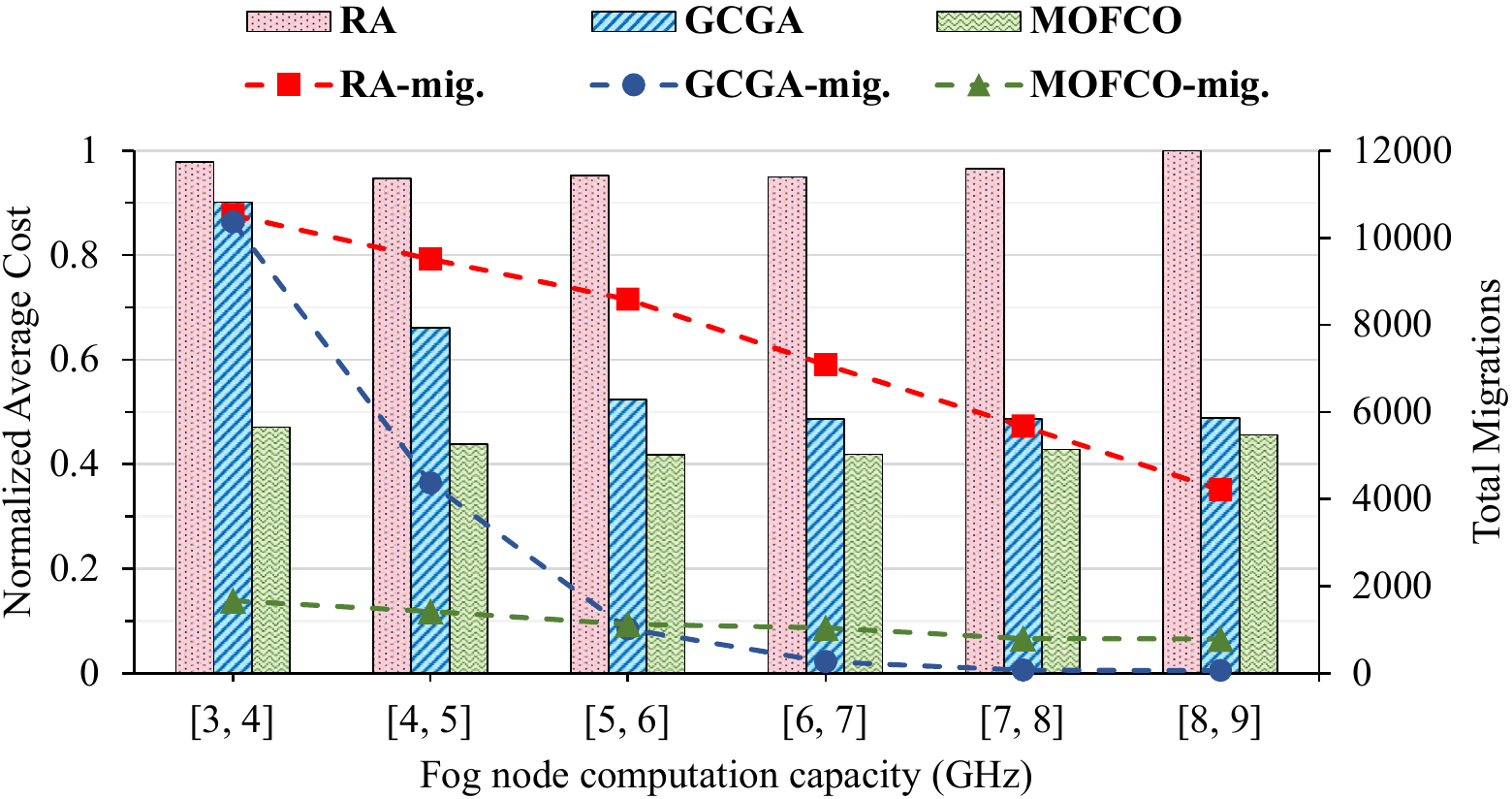}}
\caption{Impact of fog node capacities on the results. The simulation results were calculated with 70 UEs and 35 fog nodes over a timespan of 30 minutes.
}
\label{fogCapacity}
\vspace{-1.75em}
\end{figure}

For further justification, Fig.~\ref{fogCapacity} shows the effect of fog computing power. Increasing the computational power provides more computing resources but also increases energy consumption due to higher CPU frequencies. Previous methods like GCGA make offloading decisions heuristically, based on a greedy approach that avoids migrations. As seen in the figure, this results in an extremely low number of migrations once fog node capacities exceed 5\,GHz. However, higher computing power leads to increased energy consumption, and the overall cost may not be favorable. This is why MOFCO achieves lower cost compared to GCGA, despite having a higher number of migrations in some scenarios. The overall decrease in both cost and total migrations is noticeable in GCGA and MOFCO, which are capable of utilizing the fog computing resources. In these scenarios, MOFCO performs on average 15\% better than GCGA and achieves up to 43\% improvement in certain cases.
In Fig.~\ref{delta}, the effect of the constant parameter $\delta$, which represents the migration coefficient, is evaluated. A higher value of $\delta$ means that each migration imposes a higher cost on the system. It is important to note that the normalized values in this case are calculated based on the highest cost in each scenario, which corresponds to the cost of RA. GCGA reduces the number of migrations with high sensitivity to migration cost, decreasing it to less than half in the second scenario. The RA algorithm results in the highest cost and does not respond to changes in the migration coefficient. MOFCO, due to its fog node evaluation strategy, keeps the number of migrations low in all scenarios, regardless of the migration cost. This is because it relies more on mobility and location prediction, as well as workload prediction, unlike previous methods such as GCGA, which make decisions based on a greedy cost comparison. In these scenarios, MOFCO performs on average 18\% better than GCGA and achieves up to 33\% improvement in certain cases.

\section{Conclusion}

In this paper, we monitored the mobility of a certain number of UEs using a real-world SUMO simulation over a given timespan and determined the task offloading for each UE to reduce the total number of migrations. MOFCO was proposed to address the task offloading and resource allocation problem. The results show that our proposed method performs 19\% more efficient on average and can reduce the total cost by up to 43\% in certain scenarios. In future work, we aim to model the migration process more accurately, reduce its cost, and exploit it to further minimize the overall cost of execution by predicting inevitable migration situations and offloading preemptively to the correct fog node.


%

\appendices




\ifCLASSOPTIONcaptionsoff
  \newpage
\fi




\begin{thebibliography}{00}

\bibitem{b1}
S. D. Okegbile, B. T. Maharaj and A. S. Alfa, "A Multi-User Tasks Offloading Scheme for Integrated Edge-Fog-Cloud Computing Environments," in IEEE Transactions on Vehicular Technology, vol. 71, no. 7, pp. 7487-7502, July 2022, doi: 10.1109/TVT.2022.3167892.
\bibitem{b2}
P. Habibi, M. Farhoudi, S. Kazemian, S. Khorsandi and A. Leon-Garcia, "Fog Computing: A Comprehensive Architectural Survey," in IEEE Access, vol. 8, pp. 69105-69133, 2020, doi: 10.1109/ACCESS.2020.2983253.
\bibitem{b3}
H. Tran-Dang and D. -S. Kim, "FRATO: Fog Resource Based Adaptive Task Offloading for Delay-Minimizing IoT Service Provisioning," in IEEE Transactions on Parallel and Distributed Systems, vol. 32, no. 10, pp. 2491-2508, 1 Oct. 2021, doi: 10.1109/TPDS.2021.3067654.
\bibitem{b4}
H. Tran-Dang and D. -S. Kim, "Dynamic collaborative task offloading for delay minimization in the heterogeneous fog computing systems," in Journal of Communications and Networks, vol. 25, no. 2, pp. 244-252, April 2023, doi: 10.23919/JCN.2023.000008.
\bibitem{b5}
K. Wang, Y. Zhou, J. Li, L. Shi, W. Chen and L. Hanzo, "Energy-Efficient Task Offloading in Massive MIMO-Aided Multi-Pair Fog-Computing Networks," in IEEE Transactions on Communications, vol. 69, no. 4, pp. 2123-2137, April 2021, doi: 10.1109/TCOMM.2020.3046265.
\bibitem{b6}
X. Dai et al., "Task Offloading for Cloud-Assisted Fog Computing With Dynamic Service Caching in Enterprise Management Systems," in IEEE Transactions on Industrial Informatics, vol. 19, no. 1, pp. 662-672, Jan. 2023, doi: 10.1109/TII.2022.3186641.
\bibitem{b7}
Q. Wu, S. Wang, H. Ge, P. Fan, Q. Fan and K. B. Letaief, "Delay-Sensitive Task Offloading in Vehicular Fog Computing-Assisted Platoons," in IEEE Transactions on Network and Service Management, vol. 21, no. 2, pp. 2012-2026, April 2024, doi: 10.1109/TNSM.2023.3322881.
\bibitem{b8}
E. Oustad, A. Younesi, M. Ansari, S. Safari, M. A. Soleimani, J. Henkel, and A. Ejlali,
“DIST: Distributed Learning-Based Energy-Efficient and Reliable Task Scheduling and Resource Allocation in Fog Computing,” IEEE Transactions on Services Computing, vol. 18, no. 3, pp. 1336–1351, May–Jun. 2025, doi: 10.1109/TSC.2025.3568255.
\bibitem{b9}
R. A. Addad, D. L. Cadette Dutra, M. Bagaa, T. Taleb and H. Flinck, "Towards a Fast Service Migration in 5G," 2018 IEEE Conference on Standards for Communications and Networking (CSCN), Paris, France, 2018, pp. 1-6, doi: 10.1109/CSCN.2018.8581836.
\bibitem{b10}
A. Machen, S. Wang, K. K. Leung, B. J. Ko and T. Salonidis, "Live Service Migration in Mobile Edge Clouds," in IEEE Wireless Communications, vol. 25, no. 1, pp. 140-147, February 2018, doi: 10.1109/MWC.2017.1700011.
\bibitem{b11}
P. A. Lopez et al., "Microscopic Traffic Simulation using SUMO," 2018 21st International Conference on Intelligent Transportation Systems (ITSC), Maui, HI, USA, 2018, pp. 2575-2582, doi: 10.1109/ITSC.2018.8569938.
\bibitem{b12}
L. Liu, Z. Chang and X. Guo, "Socially Aware Dynamic Computation Offloading Scheme for Fog Computing System With Energy Harvesting Devices," in IEEE Internet of Things Journal, vol. 5, no. 3, pp. 1869-1879, June 2018, doi: 10.1109/JIOT.2018.2816682.
\bibitem{b13}
A. Bozorgchenani, D. Tarchi and G. E. Corazza, "Centralized and Distributed Architectures for Energy and Delay Efficient Fog Network-Based Edge Computing Services," in IEEE Transactions on Green Communications and Networking, vol. 3, no. 1, pp. 250-263, March 2019, doi: 10.1109/TGCN.2018.2885443.
\bibitem{b14}
J. Du, L. Zhao, J. Feng and X. Chu, "Computation Offloading and Resource Allocation in Mixed Fog/Cloud Computing Systems With Min-Max Fairness Guarantee," in IEEE Transactions on Communications, vol. 66, no. 4, pp. 1594-1608, April 2018, doi: 10.1109/TCOMM.2017.2787700.
\bibitem{b15}
J. Du, L. Zhao, X. Chu, F. R. Yu, J. Feng and C. -L. I, "Enabling Low-Latency Applications in LTE-A Based Mixed Fog/Cloud Computing Systems," in IEEE Transactions on Vehicular Technology, vol. 68, no. 2, pp. 1757-1771, Feb. 2019, doi: 10.1109/TVT.2018.2882991.
\bibitem{b16}
H. Shah-Mansouri and V. W. S. Wong, "Hierarchical Fog-Cloud Computing for IoT Systems: A Computation Offloading Game," in IEEE Internet of Things Journal, vol. 5, no. 4, pp. 3246-3257, Aug. 2018, doi: 10.1109/JIOT.2018.2838022.
\bibitem{b17}
M. Liu, F. R. Yu, Y. Teng, V. C. M. Leung and M. Song, "Distributed Resource Allocation in Blockchain-Based Video Streaming Systems With Mobile Edge Computing," in IEEE Transactions on Wireless Communications, vol. 18, no. 1, pp. 695-708, Jan. 2019, doi: 10.1109/TWC.2018.2885266.
\bibitem{b18}
H. Ye, G. Y. Li and B. -H. F. Juang, "Deep Reinforcement Learning Based Resource Allocation for V2V Communications," in IEEE Transactions on Vehicular Technology, vol. 68, no. 4, pp. 3163-3173, April 2019, doi: 10.1109/TVT.2019.2897134.

\begin{figure}
\centerline{\includegraphics[width=\columnwidth, right]{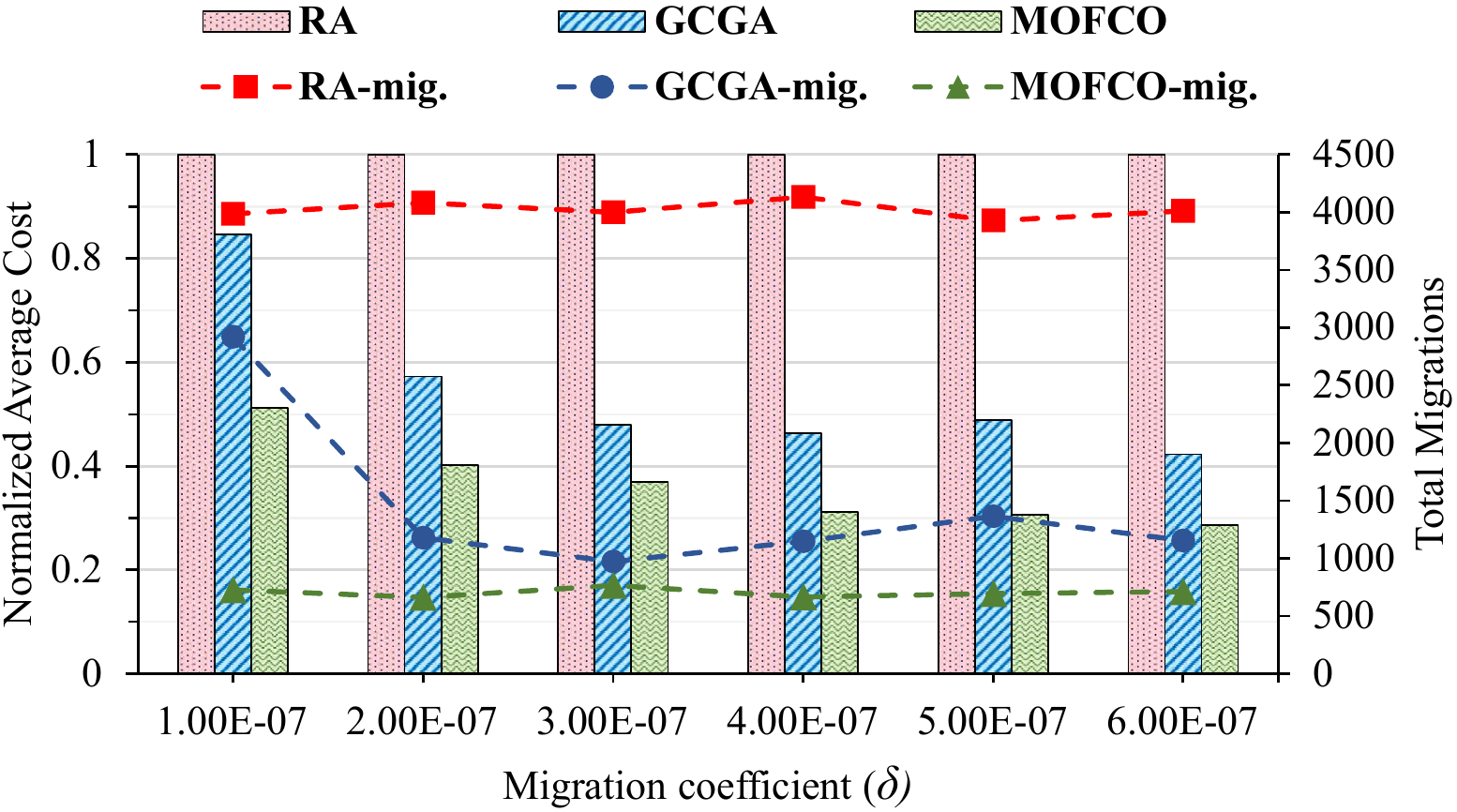}}
\caption{Impact of $\delta$ on the results. The simulation results were calculated with 70 UEs and 35 fog nodes over a timespan of 30 minutes.
}
\label{delta}
\vspace{-1.75em}
\end{figure}


\bibitem{b19}
F. Chai, Q. Zhang, H. Yao, X. Xin, R. Gao and M. Guizani, "Joint Multi-Task Offloading and Resource Allocation for Mobile Edge Computing Systems in Satellite IoT," in IEEE Transactions on Vehicular Technology, vol. 72, no. 6, pp. 7783-7795, June 2023, doi: 10.1109/TVT.2023.3238771.
\bibitem{b20}
T. X. Tran and D. Pompili, "Joint Task Offloading and Resource Allocation for Multi-Server Mobile-Edge Computing Networks," in IEEE Transactions on Vehicular Technology, vol. 68, no. 1, pp. 856-868, Jan. 2019, doi: 10.1109/TVT.2018.2881191.
\bibitem{b21}
Y. Shi, S. Chen and X. Xu, "MAGA: A Mobility-Aware Computation Offloading Decision for Distributed Mobile Cloud Computing," in IEEE Internet of Things Journal, vol. 5, no. 1, pp. 164-174, Feb. 2018, doi: 10.1109/JIOT.2017.2776252.
\bibitem{b22}
S. -S. Lee and S. Lee, "Resource Allocation for Vehicular Fog Computing Using Reinforcement Learning Combined With Heuristic Information," in IEEE Internet of Things Journal, vol. 7, no. 10, pp. 10450-10464, Oct. 2020, doi: 10.1109/JIOT.2020.2996213.
\bibitem{b23}
Z. Zhang, Z. Chen, Y. Shen, X. Dong, and N. Xi, "A Dynamic Task Offloading Scheme Based on Location Forecasting for Mobile Intelligent Vehicles," IEEE Trans. Veh. Technol., vol. PP, pp. 1–15, 2024, doi: 10.1109/TVT.2024.3351224.
\bibitem{b24}
C. Chen, Y. Zeng, H. Li, Y. Liu and S. Wan, "A Multihop Task Offloading Decision Model in MEC-Enabled Internet of Vehicles," in IEEE Internet of Things Journal, vol. 10, no. 4, pp. 3215-3230, 15 Feb.15, 2023, doi: 10.1109/JIOT.2022.3143529.
\bibitem{b25}
L. Liu, M. Zhao, M. Yu, M. A. Jan, D. Lan and A. Taherkordi, "Mobility-Aware Multi-Hop Task Offloading for Autonomous Driving in Vehicular Edge Computing and Networks," in IEEE Transactions on Intelligent Transportation Systems, vol. 24, no. 2, pp. 2169-2182, Feb. 2023, doi: 10.1109/TITS.2022.3142566.
\bibitem{b26}
Z. Zhang and F. Zeng, "Efficient Task Allocation for Computation Offloading in Vehicular Edge Computing," in IEEE Internet of Things Journal, vol. 10, no. 6, pp. 5595-5606, 15 March15, 2023, doi: 10.1109/JIOT.2022.3222408.
\bibitem{b27}
L. Zhao et al., "MESON: A Mobility-Aware Dependent Task Offloading Scheme for Urban Vehicular Edge Computing," in IEEE Transactions on Mobile Computing, vol. 23, no. 5, pp. 4259-4272, May 2024, doi: 10.1109/TMC.2023.3289611.
\bibitem{b28}
C. Zhu et al., "Folo: Latency and Quality Optimized Task Allocation in Vehicular Fog Computing," in IEEE Internet of Things Journal, vol. 6, no. 3, pp. 4150-4161, June 2019, doi: 10.1109/JIOT.2018.2875520.
\bibitem{b29}
D. Wang, Z. Liu, X. Wang and Y. Lan, "Mobility-Aware Task Offloading and Migration Schemes in Fog Computing Networks," in IEEE Access, vol. 7, pp. 43356-43368, 2019, doi: 10.1109/ACCESS.2019.2908263.
\bibitem{b30}
Y. Wen, W. Zhang and H. Luo, ”Energy-optimal mobile application execution: Taming resource-poor mobile devices with cloud clones,” in 2012 Proceedings IEEE International Conference on Computer Communications (INFOCOM), Orlando, FL, 2012, pp. 2716-2720.
\bibitem{b31}
W. Hao and S. Yang, ”Small Cell Cluster-Based Resource Allocation for Wireless Backhaul in Two-Tier Heterogeneous Networks With Massive MIMO,” IEEE Transactions on Vehicular Technology, vol. 67, no. 1, pp. 509-523, Jan. 2018.
\bibitem{b32}
R. Yadav, W. Zhang, O. Kaiwartya, H. Song and S. Yu, "Energy-Latency Tradeoff for Dynamic Computation Offloading in Vehicular Fog Computing," in IEEE Transactions on Vehicular Technology, vol. 69, no. 12, pp. 14198-14211, Dec. 2020, doi: 10.1109/TVT.2020.3040596.
\bibitem{b33}
C. Tang, X. Wei, C. Zhu, Y. Wang and W. Jia, "Mobile Vehicles as Fog Nodes for Latency Optimization in Smart Cities," in IEEE Transactions on Vehicular Technology, vol. 69, no. 9, pp. 9364-9375, Sept. 2020, doi: 10.1109/TVT.2020.2970763.
\bibitem{b34}
H. Escobar-Cuevas, E. Cuevas, A. Luque-Chang, O. Barba-Toscano, and M. Pérez-Cisneros, "Enhancing metaheuristic algorithm performance through structured population and evolutionary game theory," in Mathematics, vol. 12, no. 23, p. 3676, 2024, doi.org/10.3390/math12233676.
\bibitem{b35}
W. Nasrin and J. Xie, "SharedMEC: Sharing Clouds to Support User Mobility in Mobile Edge Computing," 2018 IEEE International Conference on Communications (ICC), Kansas City, MO, USA, 2018, pp. 1-6, doi: 10.1109/ICC.2018.8422241.
\bibitem{b36}
DLR-TS, "TAVF-Hamburg," GitHub. [Online]. Available: https://github.com/DLR-TS/sumo-scenarios/tree/main/TAVF-Hamburg. [Accessed: Sep. 23, 2024].

\newpage

\end{thebibliography}
%

%

\begin{IEEEbiography}
[{\includegraphics[width=1in,height=1.25in,clip,keepaspectratio]{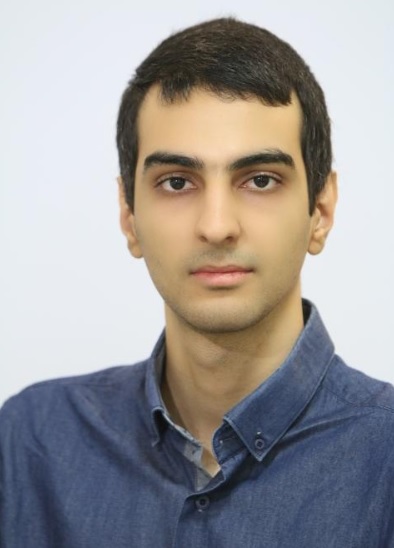}}]{Soheil Mahdizadeh} received his B.Sc. degree in Computer Engineering from Sharif University of Technology, Tehran, Iran, in 2023, where he was a top-ranked student. He is currently pursuing his M.Sc. degree at the same university and is a member of Sharif's Cyber-Physical Systems Laboratory. His research interests include fog computing, mobile fog computing, game theory, internet of things, embedded systems, cyber-physical systems, and internet of vehicles.
\end{IEEEbiography}



\begin{IEEEbiography}[{\includegraphics[width=1in,height=1.25in,clip,keepaspectratio]{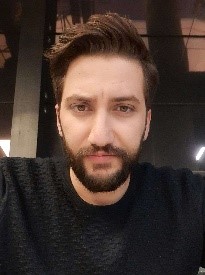}}]{Elyas Oustad}
is currently a PhD student in computer engineering at Sharif University of Technology, Tehran, Iran. Received his MSc degree in computer engineering from the Iran University of Science and Technology, Tehran, Iran, in 2022. He is currently a member of the Cyber-Physical Systems Laboratory at Sharif University of Technology. He was a Cyber-Physical Systems Laboratory member at Iran University of Science and Technology from 2020 to 2022. His research interests include  internet of things, embedded systems, cyber-physical systems, fog computing, internet of vehicles, and internet of medical things.
\end{IEEEbiography}

\begin{IEEEbiography}[{\includegraphics[width=1in,height=1.25in,clip,keepaspectratio]{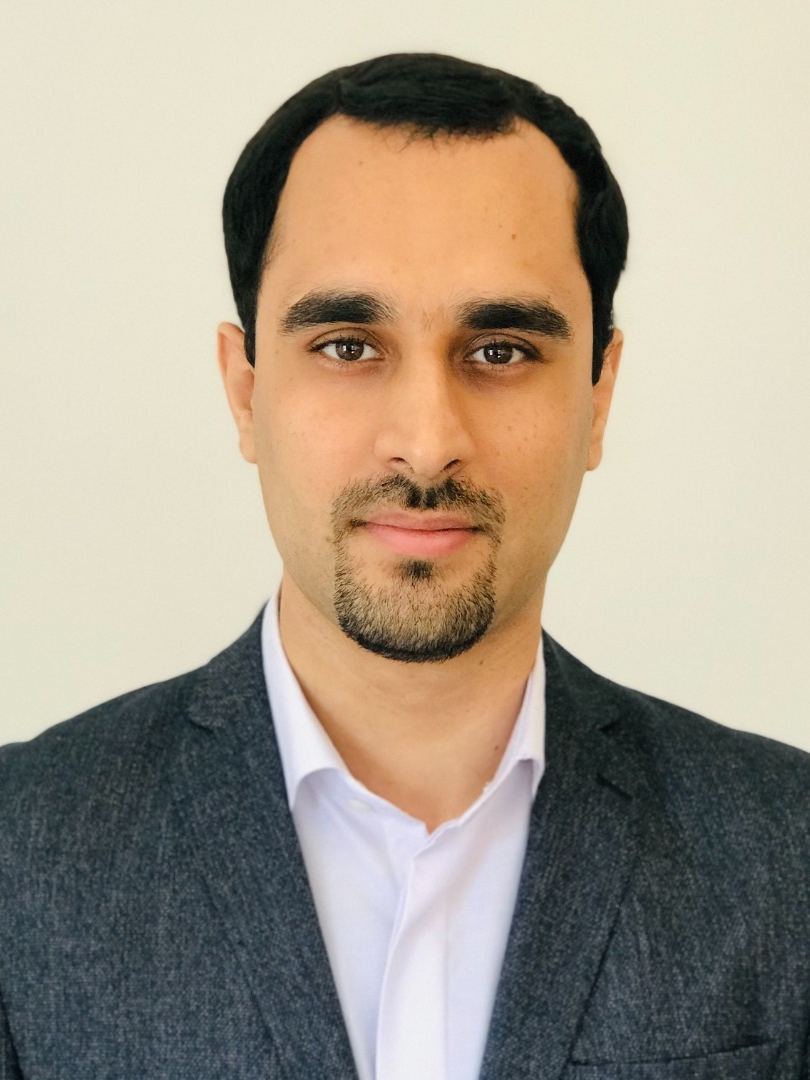}}]
{Mohsen Ansari} is currently an assistant professor of computer engineering at Sharif University of Technology, Tehran, Iran. He received his Ph.D. degree in computer engineering from Sharif University of Technology, Tehran, Iran, in 2021. He was a visiting researcher in the Chair for Embedded Systems (CES), Karlsruhe Institute of Technology (KIT), Germany, from 2019 to 2021. He is currently the director of the Cyber-Physical Systems Laboratory (CPSLab) at Sharif University of Technology. He was the technical program committee (TPC) member of ASP-DAC (2022, 2023, and 2024). Dr. Ansari is serving as an associate editor of the IEEE Embedded Systems Letter (ESL). His research interests include embedded machine learning, edge/fog/cloud computing, low-power design, real-time systems, cyber-physical systems, and hybrid systems design.
\end{IEEEbiography}




\end{document}